\documentclass[twocolumn]{aastex63}

\submitjournal{ApJ}

\shorttitle{Atmospheric dynamics on eccentric orbits}
\shortauthors{I. Guendelman \& Y.Kaspi}
\graphicspath{{./}{Figs/}}
\begin{document}

\title{Atmospheric dynamics on terrestrial planets with eccentric orbits}

\correspondingauthor{Ilai Guendelman}
\email{ilai.guendelman@weizmann.ac.il}

\author[0000-0002-6873-0320]{Ilai Guendelman}
\affiliation{Department of Earth and Planetary Sciences, Weizmann Institute of \
Science \\
234 Herzl st., 76100 \\
Rehovot, Israel}

\author[0000-0003-4089-0020]{Yohai Kaspi}
\affiliation{Department of Earth and Planetary Sciences, Weizmann Institute of \
Science \\
234 Herzl st., 76100 \\
Rehovot, Israel}



\begin{abstract}
The insolation a planet receives from its parent star is the main driver of the climate and depends on the planet's orbital configuration. Planets with non-zero obliquity and eccentricity experience seasonal insolation variations. As a result, the climate exhibits a seasonal cycle, with its strength depending on the orbital configuration and atmospheric characteristics. In this study, using an idealized general circulation model, we examine the climate response to changes in eccentricity for both zero and non-zero obliquity planets. In the zero obliquity case, a comparison between the seasonal response to changes in eccentricity and perpetual changes in the solar constant shows that the seasonal response strongly depends on the orbital period and radiative timescale. More specifically, using a simple energy balance model, we show the importance of the latitudinal structure of the radiative timescale in the climate response. We also show that the response strongly depends on the atmospheric moisture content. The combination of an eccentric orbit with non-zero obliquity is complex, as the insolation also depends on the perihelion position. Although the detailed response of the climate to variations in eccentricity, obliquity, and perihelion is involved, the circulation is constrained mainly by the thermal Rossby number and the maximum temperature latitude. Finally, we discuss the importance of different planetary parameters that affect the climate response to orbital configuration variations.
\end{abstract}



\section{Introduction} \label{sec:intro}
\begin{figure}[ht!]
\centering
\includegraphics[width=\columnwidth]{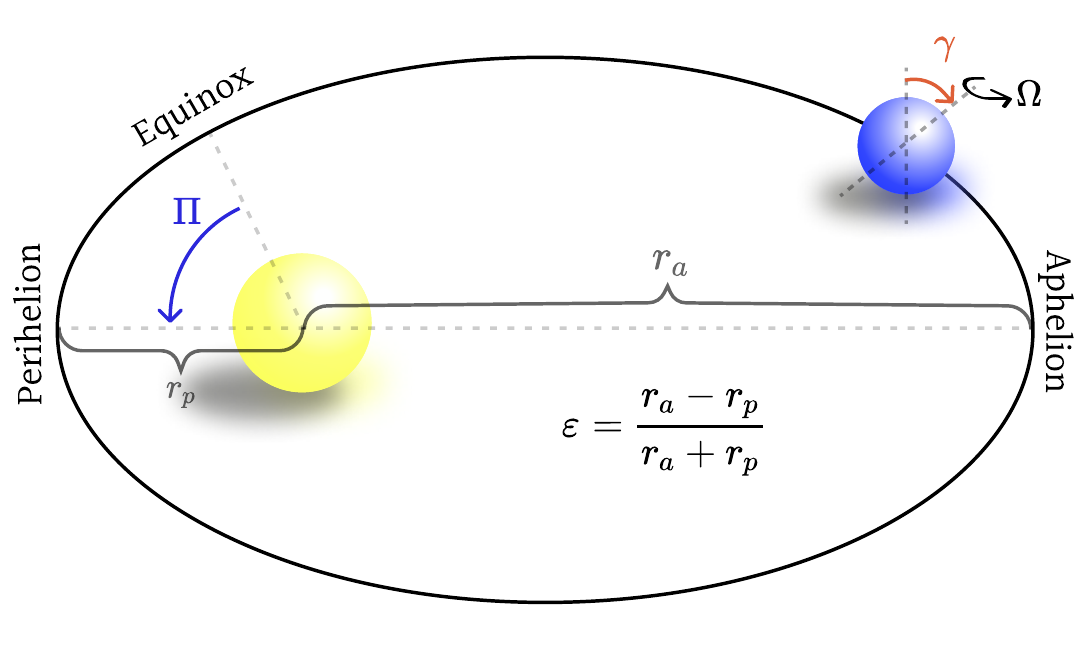}
\caption{Schematic plot of a planet's orbit and the relevant parameters, obliquity ($\gamma$), eccentricity ($\varepsilon$), and perihelion ($\Pi$). $r_p$ and $r_a$ are the distances from the star at perihelion and aphelion, respectively, and $\Omega$ is the rotation rate. \label{fig:orbit1}}
\end{figure}

The climate on a planetary body is sensitive to the planet's characteristics \citep[e.g.,][]{kaspi2015,komacek2019}. In particular, the planet's orbital configuration significantly affects the climate system, as it dictates the incoming solar radiation. More specifically, the orbital configuration, namely, the obliquity ($\gamma$), eccentricity ($\varepsilon$), and perihelion ($\Pi$) dictate the insolation seasonal cycle (Fig.~\ref{fig:orbit1} depicts a schematic plot of the orbit and the orbital parameters). In addition to the reasonable assumption that a wide set of orbital configurations exists across the universe, the orbital configuration of the different planets changes with a Milankovitch-like cycle \citep[e.g.,][]{spiegel_generalized_2010}. This poses the question of how the atmospheric dynamics depends on the orbital configuration.

\begin{figure}[ht!]
\centering
\includegraphics[width=\columnwidth]{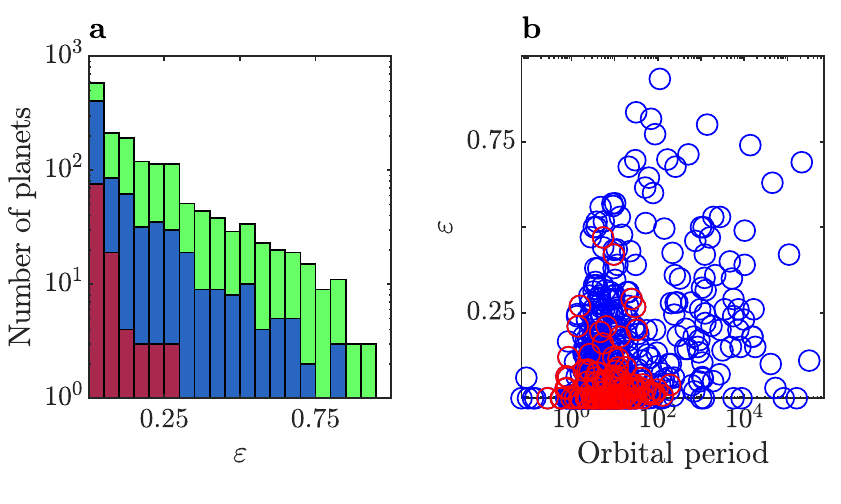}
\caption{a) Histogram of eccentricity values of the confirmed exoplanets; data taken from the \href{http://www.exoplanet.eu}{exoplanet.eu} catalog. Shown are all planets with measured eccentricity (green), all the planets with an observed mass (blue) and all the planets with an observed mass less than 10 times the mass of Earth (red). b) Scatter plot of all the planets with a measured mass, eccentricity, and orbital period (in days); shown are all the planets with a measured orbital period and eccentricity (blue) and all the planets with a mass less than 10 times the mass of Earth (red). \label{fig:distri}}
\end{figure}

As eccentricity is a measurable quantity for some confirmed exoplanets, one can look at the measured eccentricity distribution, which demonstrates that it spans all eccentricity values (Fig.~\ref{fig:distri}). However, low-mass planets, i.e., planets with a mass lower than $10$ times the mass of Earth (more relevant for this study), do not span the entire range of eccentricities, with Kepler-68c having the largest eccentricity value \citep[$\varepsilon=0.42$,][]{gilliland2013kepler}. On the one hand, small mass planets may be prone to have small eccentricities \citep{howard2013observed}, while, on the other hand, it seems that most of the observed low mass planets are in a close-in orbit (Fig.~\ref{fig:distri}), and that future observations will reveal the existence of more eccentric low-mass planets. Nonetheless, current observations suggest that eccentricity varies within a significant range, motivating the question of how atmospheric dynamics depend on eccentricity.

\begin{figure*}[htb!]
\centering
\includegraphics[width=0.8\paperwidth]{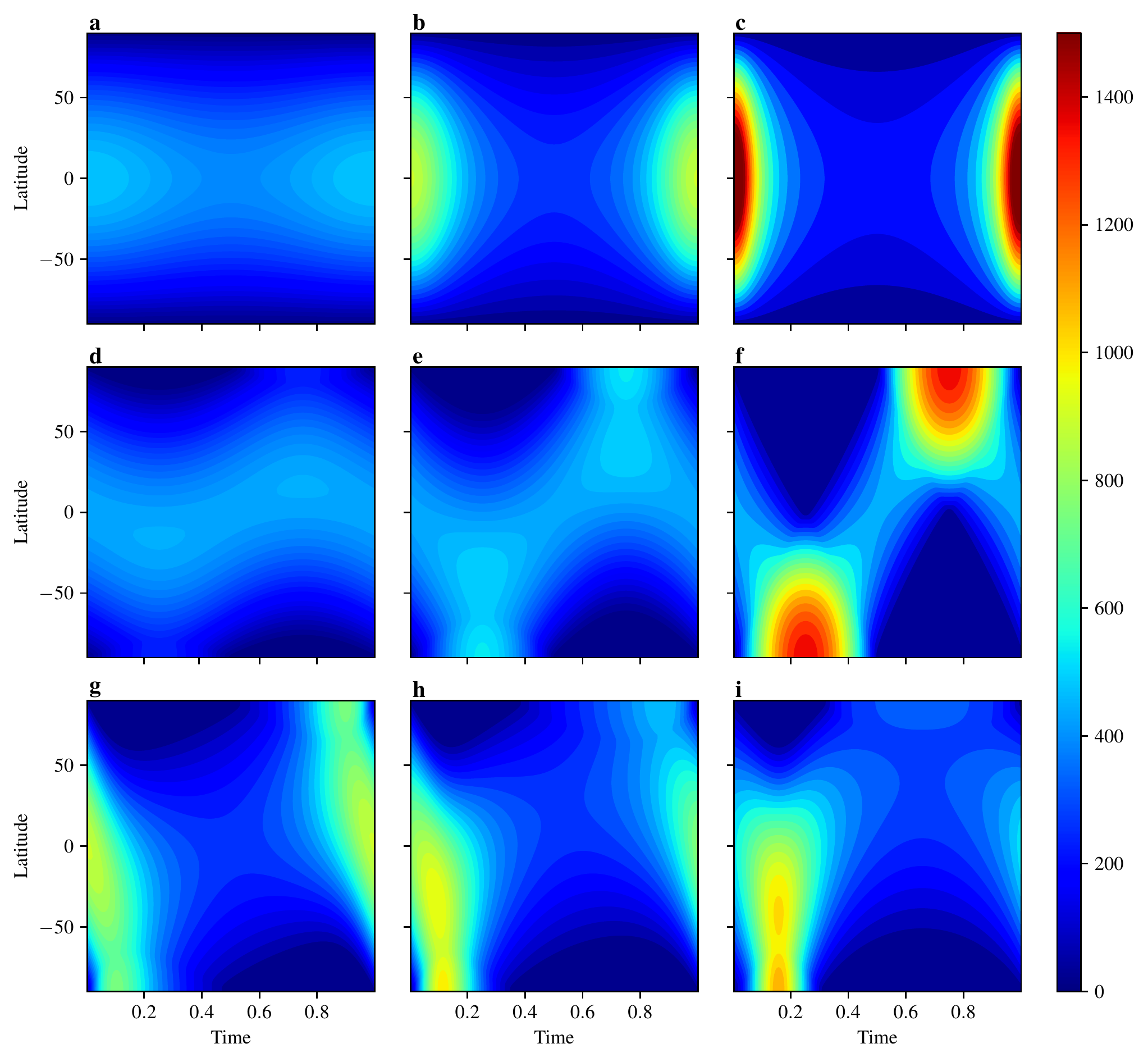}
\caption{Diurnal mean insolation (time unites are normalized) for different values of eccentricity, obliquity and perihelion. Panels a-c are zero obliquity cases with $\epsilon=0.05,0.3$ and $0.5$, respectively. Panels d-f are zero eccentricity cases with $\gamma=10^{\circ}, 23^{\circ}$ and $90^{\circ}$, respectively, with perihelion at $0^{\circ}$. Panels g-i are for $\epsilon=0.3$ and $\gamma=23^{\circ}$ with perihelion $0^{\circ}$, $45^{\circ}$ and $90^{\circ}$, respectively. \label{fig:inso1}}
\end{figure*}

Each of the mentioned parameters ($\gamma$, $\varepsilon$ and $\Pi$) adds a seasonal cycle of a different nature to the insolation, non-zero obliquity introduces seasonal variations in the latitudinal insolation structure (Fig.~\ref{fig:inso1}d-f). The eccentricity seasonal cycle is effectively a seasonal cycle of the solar constant (Fig.~\ref{fig:inso1}a-c). The perihelion position becomes important in planets with non-zero obliquity and eccentricity, where the phase between the closest approach (perihelion, higher solar constant) and equinox becomes relevant (see Fig.~\ref{fig:inso1}g-i; a more detailed discussion about this point is given in section~\ref{sec:obliq_ecc}).

The atmospheric response to the seasonally varying insolation depends on different planetary and atmospheric characteristics, specifically, the orbital period and the atmospheric radiative timescale. Longer orbital periods give the atmosphere more time to adjust to the insolation seasonal cycle, resulting in a stronger seasonal cycle. A longer radiative timescale translates into a weaker seasonal cycle, as the atmosphere needs more time to adjust to changes in the radiation \citep{Mitchell_2014, guendelman2019}.

The effect of eccentricity varies depending on the planet's orbital configuration. It is useful to distinguish between three configurations. The first, a tidally locked configuration; in this case, in addition to the variations in the solar constant during the orbital period, on eccentric tidally locked planets, the rotation rate is pseudo-synchronized, such that the rotation rate is synchronized at perihelion, and due to variations of the orbital velocity during the orbital period, the sub-stellar longitude is liberated \citep{hut1981tidal}. Numerous studies have been carried out regarding the effect of eccentricity on the habitability and atmospheric dynamics of tidally locked planets \citep[e.g.,][]{lewis_atmospheric_2010, kataria_three_2013, wang_climate_2014, lewis_atmospheric_2014, bolmont_habitability_2016}. Among them, \citet{kataria_three_2013} studied the atmospheric dynamics of a tidally locked planet with eccentricity, taking pseudo-synchronization into account. They showed that, over a large range of eccentricities, the circulation characteristics stay similar to a circular tidally locked orbit, and that the seasonal changes are mostly quantitative in nature. More recently, \citet{lewis_atmospheric_2017} have studied the extreme case of HD 80606b ($\varepsilon=0.93$); in this extreme case, when considering pseudo-synchronization, the circulation shifts during the orbital period from a tidally locked climate to a more diurnal mean, zonally symmetric one.

The second and third configurations are for planets where the diurnal mean insolation is the dominant forcing, similar to Earth's case. The difference between the two lies in their obliquity: for one, the obliquity is zero, and for the other, the obliquity is non-zero. For both these cases, previous studies have focused mainly on how eccentricity affects the planetary habitability and on the transition to a snowball state \citep[e.g.,][]{williams_earth-like_2002, dressing_habitable_2010, spiegel_generalized_2010, linsenmeier_climate_2015, mendez_equilibrium_2017}. The methods used in those studies range from energy balance models \citep[EBM, e.g.,][]{dressing_habitable_2010} to simple hydrodynamical models \citep[e.g.,][]{adams_aquaplanet_2019, ohno_atmospheres_2019} and comprehensive general circulation models \citep[GCM, e.g.,][]{williams_earth-like_2002, way_effects_2017}. \citet{ohno_atmospheres_2019}, using a simple 1.5 layer model, studied the climate response to different orbital forcing and radiative timescale, showing that depending on the specific orbital configuration and radiative timescale, the climate shifts from being forced by the annual mean forcing (mainly in relatively long radiative timescales) to being forced by seasonal forcing (in relatively short radiative timescales). In addition, they showed that in some configurations during the seasonal cycle, there is a transition from a climate that is controlled by the diurnal mean to one governed by the diurnal cycle.

In this study, we use an idealized GCM to systematically study the effect of eccentricity on the climate. For simplicity, we focus on the diurnal mean forcing and explore eccentricity values up to $0.5$. The simplest configuration of a seasonal cycle due to eccentricity is a planet in an eccentric orbit with zero obliquity. As the seasonal cycle on such planets is equivalent to seasonal variation in the solar constant, we start by considering the effect of changing the solar constant for a perpetual equinox  in section~\ref{sec:perp_solar}. We show that the climate response differs between dry and moist atmospheres, a result of the nonlinear response of moisture to changes in temperature. The perpetual equinox case acts as a baseline for the study of the seasonal cycle on planets in an eccentric orbit with zero eccentricity in section~\ref{sec:zero_obliq}. In section~\ref{sec:obliq_ecc}, we present the complexity that arises when combining eccentricity and obliquity, giving constraints on the circulation response and discussing the important parameters in this problem. Finally, we conclude in section~\ref{sec:conc}.

\section{Perpetual solar constant variations} \label{sec:perp_solar}
The few studies concerning the eccentricity effect on planets with zero obliquity have focused mainly on the temperature response and less on the atmospheric dynamics \citep[e.g.,][]{dressing_habitable_2010, ohno_atmospheres_2019}. \citet{kane2017} compared the effect of eccentricity and obliquity on the insolation, showing that even for low eccentricity values, the effect of eccentricity is significant. Motivated by that, and for the sake of completeness, we start by examining the simpler case of zero obliquity before delving into the more complex cases.

An idealized general circulation model with a seasonal cycle \citep{guendelman2019} is used in this study. This model has a simplified moisture representation \citep{frierson2006}. For simplicity, the optical depth is taken to be constant in latitude, meaning we neglect water-vapor feedback. Although using a more complex GCM, which includes water vapor feedback, clouds, and sea-ice, might affect the results, the idealized configuration is a good starting point for the study of the climate sensitivity to eccentricity.

\begin{figure*}[ht!]
\centering
\includegraphics[width=0.8\paperwidth]{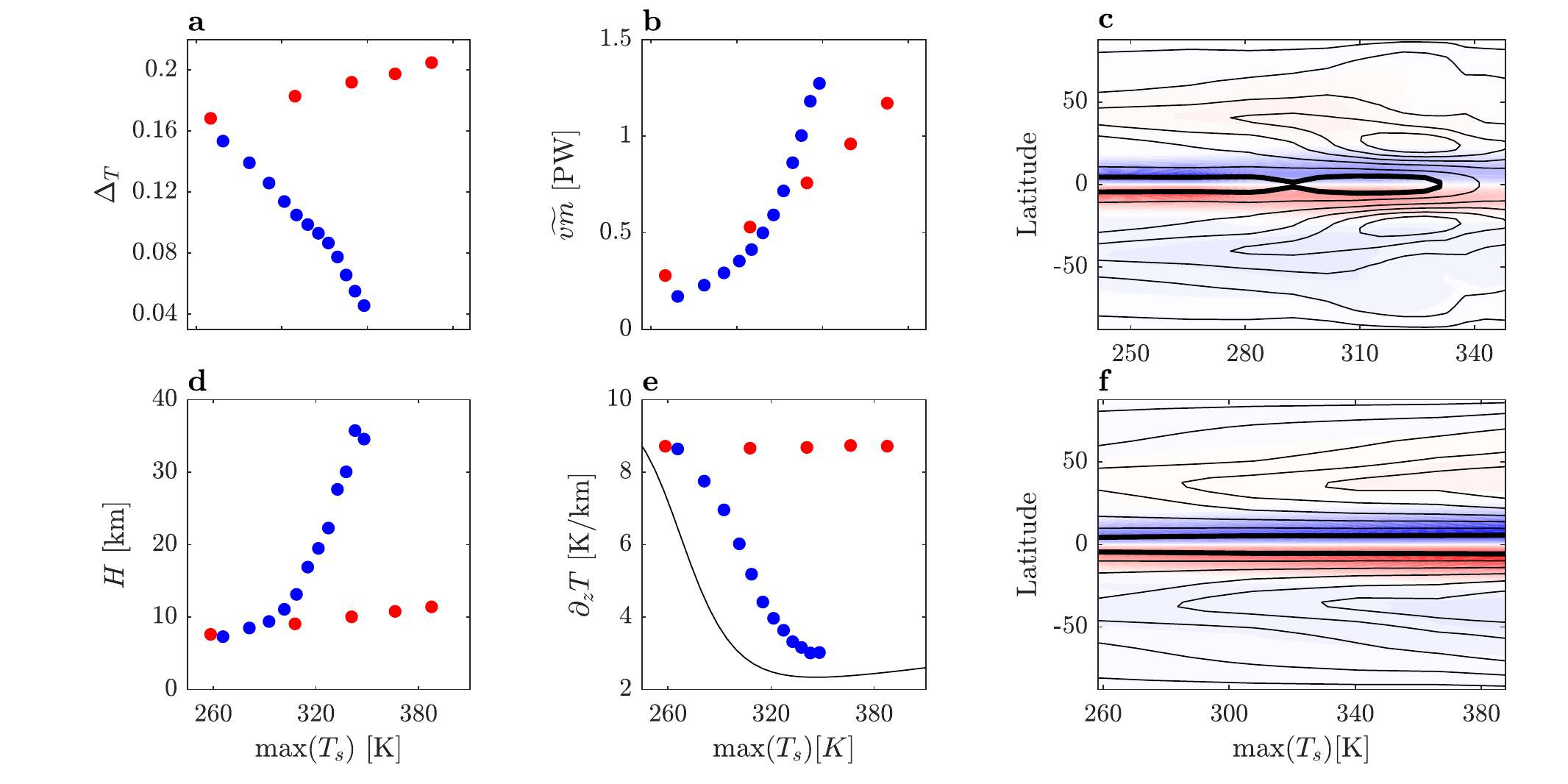}
\caption{Comparison between the moist (blue dots) and dry (red dots) simulation for increasing values of $S_0$ (corresponds to increasing values of maximum surface temperature ($\max(T_s)$). Note that for the moist simulations the range of $S_0$ is $500-3500$ $\rm{Wm^{-2}}$, and for the dry simulations the range of $S_0$ is $500-2500$ $\rm{Wm^{-2}}$.  Panels a, b, d and e are the normalized meridional temperature gradient ($\Delta_T$), Northern Hemisphere mean MSE flux ($\widetilde{vm}$, where tilde denotes a vertical average and an average over the NH of $\overline{vm}$) tropopause height ($H$) and tropospharic lapse rate ($\partial_zT$) as a function of $\max(T_s)$, respectively. Panels c and f, respectively, are the moist and dry mean meridional circulation (colors, vertically averaged between 400 and 600 mbar, blue denotes the northward flow in the upper branch of the circulation), and the zonal mean zonal wind (contours, vertically averaged between 100 and 500 mbar), where the bold contour represents the zero zonal mean zonal wind line. Note that the colorscale range differs, where in c it is $\pm1\times 10^{11} \ \rm{kg \ s^{-1}}$ and in f it is $\pm2\times10^{11} \ \rm{kg \ s^{-1}}$. The black line in panel e is the saturation moist adiabat at 600 mbar. \label{fig:solar1}}
\end{figure*}

The insolation variations during an eccentric orbit of a zero obliquity planet are equivalent to changes in the solar constant ($S_0$) during the orbit. For this reason, before focusing on the eccentricity seasonal cycle, it is beneficial to study the response of the dynamics to perpetual changes in the solar constant. Most previous studies of the climate dependence on the solar constant were done with the purpose of determining planetary habitability \citep[e.g.,][]{selsis2007, abe2011, kopparapu2013, leconte2013, shields2013, wolf2015, godolt2015, popp2016, wolf2017constraints}. In addition, studies that did focus on the atmospheric dynamics response to the solar constant variations were done as part of large parameter sweep, discussing only briefly the solar constant effect \citep[e.g.,][]{kaspi2015, komacek2019}. In addition to the trivial warming with $S_0$, \citet{kaspi2015} found that the normalized equator-to-pole temperature difference
\begin{equation}
    \Delta_T=\frac{\max(T_s)-\min(T_s)}{\rm{mean}(T_s)},
\end{equation}
where $T_s$ is the surface temperature, changes in a non-monotonic form with $S_0$, where for small $S_0$, $\Delta_T$ increases with $S_0$, and for high $S_0$, $\Delta_T$ decreases with $S_0$ (Fig.~\ref{fig:solar1}a). \citet{kaspi2015} attributed the non-monotonic behavior of $\Delta_T$ with $S_0$ to the non-linearity of moisture with temperature. This non-linearity results in more efficient equator-to-pole heat transport as the climate gets warmer (Fig.~\ref{fig:solar1}b). The total heat transport can be described in terms of the flux of moist static energy (MSE), $m=Lq+s$, where $L$ is the latent heat of vaporization, $q$ is the specific humidity, and $s=C_pT+gz$ is the dry static energy, where $C_p$ is the heat capacity of dry air, $T$ is the temperature, $g$ is the surface gravity, and $z$ is the geopotential height. The zonal mean MSE flux, $\overline{vm}$, where $v$ is the meridional wind and bar denotes zonal mean, can be divided into contributions from the zonal mean and eddies (deviations from the zonal mean, denoted by a prime, for a general field $A$, $A'=A-\overline{A}$) in the following form
\begin{equation} \label{eq:mse_flux}
    \overline{vm}=\bar{v}\bar{m}+\overline{v'm'}=L\overline{vq}+\overline{vs}=L\bar{v}\bar{q}+\bar{v}\bar{s}+L\overline{v'q'}+\overline{v's'}.
\end{equation}

\begin{figure*}[ht!]
\centering
\includegraphics[width=0.8\paperwidth]{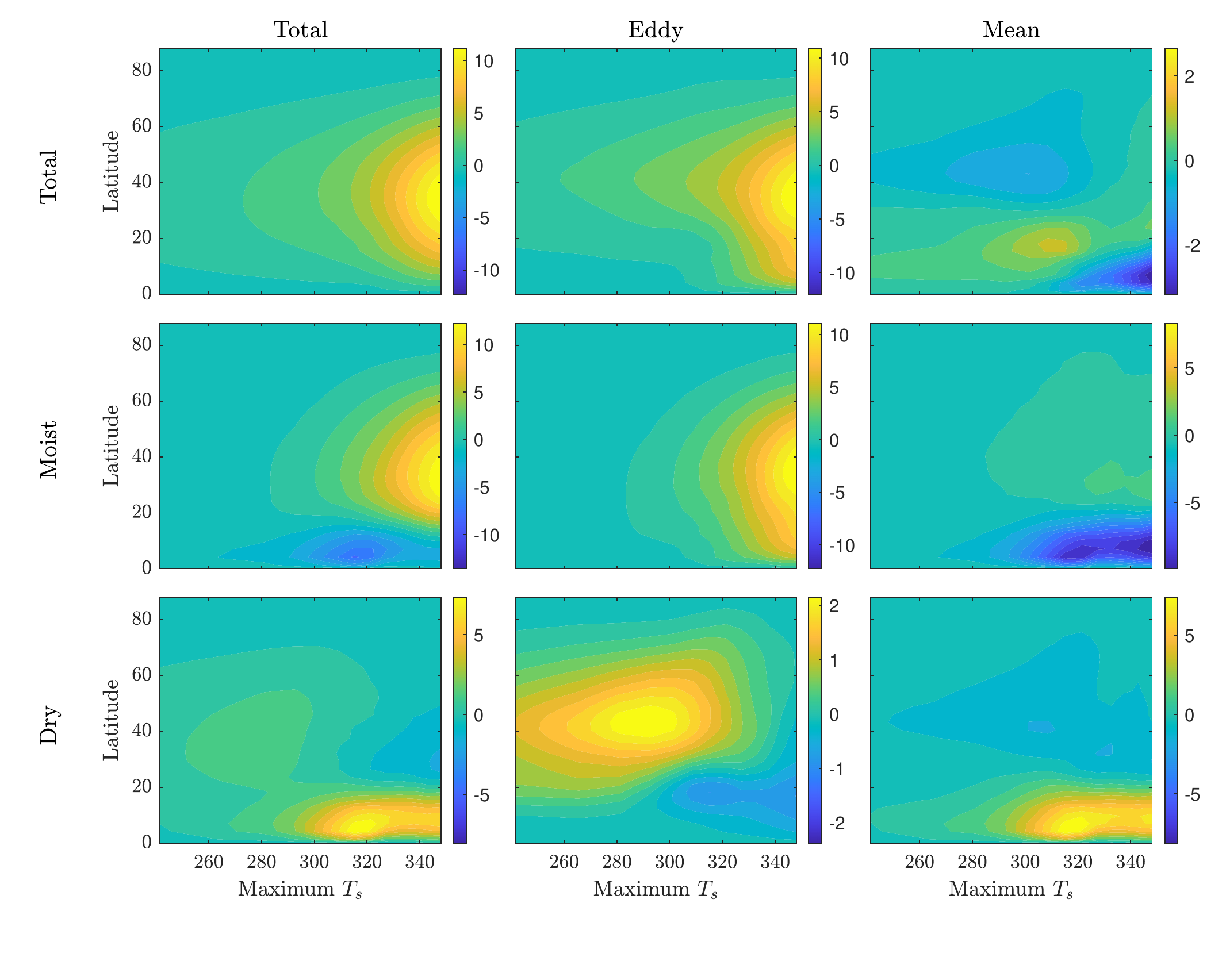}
\caption{Vertically integrated moist static energy flux and its decomposition (Eq.~\ref{eq:mse_flux}) as a function of maximum surface temperature (K). Increasing the maximum surface temperature corresponds to increasing the solar constant, raging from $500$ to $3500 \rm{ \ Wm^{-2}}$. Note that each subplot has its own colorscale (PW). \label{fig:heat_flux}}
\end{figure*}

Increasing the solar constant enhances the total heat flux, with the main contribution coming from the eddy fluxes (Fig.~\ref{fig:heat_flux}). In particular, the moist contribution becomes more dominant as the solar constant increases in a nonlinear form (Figures~\ref{fig:heat_flux} and \ref{fig:solar1}b). This non-linearity of the MSE flux explains the non-monotonic behavior of $\Delta_T$ with $S_0$ \citep{kaspi2015}. In order to illustrate this, it is convenient to look at the Clausius-Clapeyron equation for the saturation water vapor pressure in the atmosphere
\begin{equation}
    e_s(T)=e_0\exp\left[-\frac{L}{R_v}\left(\frac{1}{T}-\frac{1}{T_0}\right)\right],
\end{equation}
where $R_v$ is the gas constant for water vapor and $e_0$ is the saturation vapor pressure at $T_0=273.16^{\circ}$K. This non-linearity will lead to a higher $e_s$ in warmer latitudes. Increasing the solar constant will enhance this effect, resulting in an increased moisture meridional gradient; in order to reduce this gradient, the flux will increase, resulting in more heat transported from the equator to the poles.

To verify that the non-monotonic behavior of $\Delta_T$ with $S_0$ is a result of moisture, it is convenient to follow the approach of \citet{frierson2006}, setting $e_0$ to zero in order to eliminate moisture from the simulations. Indeed, in this 'dry' model configuration, $\Delta_T$ strictly increases with $S_0$ (Fig.~\ref{fig:solar1}a). In addition, the dry and moist simulations exhibit other significant differences. First, the dry simulations are warmer than the moist ones, a result of water evaporation acting as an energy sink (at the surface) in the moist simulations. In addition to the all-around cooling in the moist simulation, the evaporation is stronger in the equatorial regions, cooling the equator more than the poles, which results in a weaker equator-to-pole temperature difference in the moist simulations (Fig~\ref{fig:solar1}a). An additional mechanism that can explain the cooling in the moist simulations is the lapse rate feedback. The lapse rate feedback is a negative feedback, in which due to a lower lapse rate the atmosphere will radiate with higher temperatures leading to a global cooling \citep{hartmann2016}.

The moist and dry simulations also differ in the tropopause height and lapse rate (Fig.~\ref{fig:solar1}d-e). While the lapse rate in the dry simulations remains constant across the different $S_0$ values, for the moist case, it decreases with $S_0$ (Fig.~\ref{fig:solar1}e). This difference is a result of the atmosphere relaxing towards a different relevant adiabatic lapse rate. While the dry adiabat, $\Gamma_d=g/c_p$, is determined by planetary parameters that are independent of $S_0$, in the moist case, the relevant lapse rate is the saturation moist adiabatic lapse rate, which can be written as 
\begin{equation}
    \Gamma_m=\Gamma_d\frac{1+\frac{L\mu_s}{R_dT}}{1+\frac{L^2\mu_s}{C_pR_vT^2}},
\end{equation}
where $\mu_s=R_d e_s(T)/R_vp$ is the saturation mixing ratio and $R_d$ is the gas constant of dry air \citep{andrews_2010}. $\Gamma_m$ represents the lower limit of the lapse rate, which is lower than the dry adiabat and generally decreases with temperature (black line in Fig. \ref{fig:solar1}e).

The difference in the tropopause height response between the simulations can be explained by using the equation for the tropopause height \citep{vallis2015}
\begin{equation}
    H=\frac{1}{16\Gamma}\left(\lambda T_{\rm{trop}} +\sqrt{\lambda^2T_{\rm{trop}}^2 +32\Gamma\tau_sH_sT_{\rm{trop}}}\right),
\end{equation}
where $\Gamma$ is the lapse rate, $\lambda$ is a constant, $T_{\rm{trop}}$ is the topopause temperature, $\tau_s$ is the optical depth at the surface and $H_s$ is the atmospheric height scale. $H$ is proportional to $T_{\rm{trop}}$, which increases with $S_0$ in both the dry and moist cases. In addition, $H$ is inversely proportional to $\Gamma$, which decreases only in the moist case. This explains the difference in the response of $H$ between the moist and dry simulations. 

These changes in $\Delta_T, \ H,$ and lapse rate can be used to explain how the atmospheric dynamics changes as a function of $S_0$, and more specifically, the changes in the zonal mean zonal wind, $\overline{u}$, and in the mean meridional circulation, $\psi$. The mean meridional circulation is described using the mean meridional streamfunction
\begin{equation}
    \psi=\frac{2\pi a}{g}\int \overline{v}\cos\phi dp,
\end{equation}
where $a$ is the planetary radius, $\phi$ is latitude, and $p$ is pressure. On Earth, the meridional circulation is composed mainly of the tropical thermally driven Hadley cell, where, in the annual mean, air rises at the equator and descends in the subtropics. In the midlatitudes, there is the eddy-driven Ferrel cell, which is driven by baroclinic macroturbulence in the atmosphere \citep{held2000general, vallis_2017}. In both moist and dry cases, $\overline{u}$ and $\psi$ exhibit relatively small changes with $S_0$, with a general increase in the jet strength and small changes in the streamfunction with $S_0$ (Fig. \ref{fig:solar1}c,f). While the strength of the circulation increases with $S_0$ in the dry case, in the moist case it decreases (Fig.~\ref{fig:solar1}c,f). This behavior correlates with the response of $\Delta_T$ to $S_0$, and is in agreement with the axisymmetric theory that projects that the strength of the circulation is proportional to the meridional temperature gradient \citep{held1980}. This, however, should be taken with a grain of salt, as according to the \citet{held1980} scaling, the strength should increase with higher $H$; yet, it was shown that the relation between $H$ and the circulation strength is not as robust as the relation between $\Delta_T$ and the circulation strength \citep{chemke2017}.

For the moist case, simulations with high values of $S_0$, exhibit equatorial superrotation (Fig.~\ref{fig:solar1}c). A possible reason for the transition to superrotation is the decrease in $\Delta_T$ with the solar constant, which was shown to result in superrotation for some cases \citep{laraia2015, polichtchouk2016, Imamura2020}. As the equator-to-pole temperature difference decreases, so does the baroclinicity, allowing superrotation to develop from a wave source in the equatorial region \citep{polichtchouk2016}. Determining the specific mechanisms responsible for the superrotaiton transition in the simulations is beyond the scope of this study. 

\section{The seasonal cycle on a planet in an eccentric orbit with zero obliquity} \label{sec:zero_obliq}
The insolation seasonal cycle of a planet in an eccentric orbit with zero obliquity can be described as seasonal variations of the solar constant. The atmospheric response to the seasonal cycle insolation is dominated by some ratio of the radiative timescale and the orbital period \citep{Mitchell_2014, rose2017ice, guendelman2019}. Longer orbital periods give the atmosphere more time to adjust to the seasonally varying insolation, resulting in a climate with a stronger seasonal cycle and with the temperature becoming more in phase with the insolation \citep{Mitchell_2014, guendelman2019}. The radiative timescale can be written as
\begin{equation}\label{eq:rad_time}
    \tau_{\rm{rad}}=\frac{C}{4\sigma T_e^3},
\end{equation}
where $C$ is the atmospheric heat capacity, $\sigma$ is the Stefan-Boltzmann constant and $T_e$ is the equilibrium temperature, 
\begin{equation}\label{eq:eq_temp}
    T_e=\left(\frac{Q}{\sigma}\right)^{1/4},
\end{equation}
where $Q$ is the incoming insolation at the top of the atmosphere. The radiative timescale controls the time the atmosphere needs to adjust to radiative changes. Substituting Equation~\ref{eq:eq_temp} into Equation \ref{eq:rad_time} gives $\tau_{\rm{rad}}\propto Q^{-3/4}$, meaning that the radiative timescale is inversely proportional to the insolation. Alternatively, colder temperatures correspond to longer time for the atmospheric response to radiative changes.

\begin{figure*}[ht!]
\centering
\includegraphics[width=0.82\paperwidth]{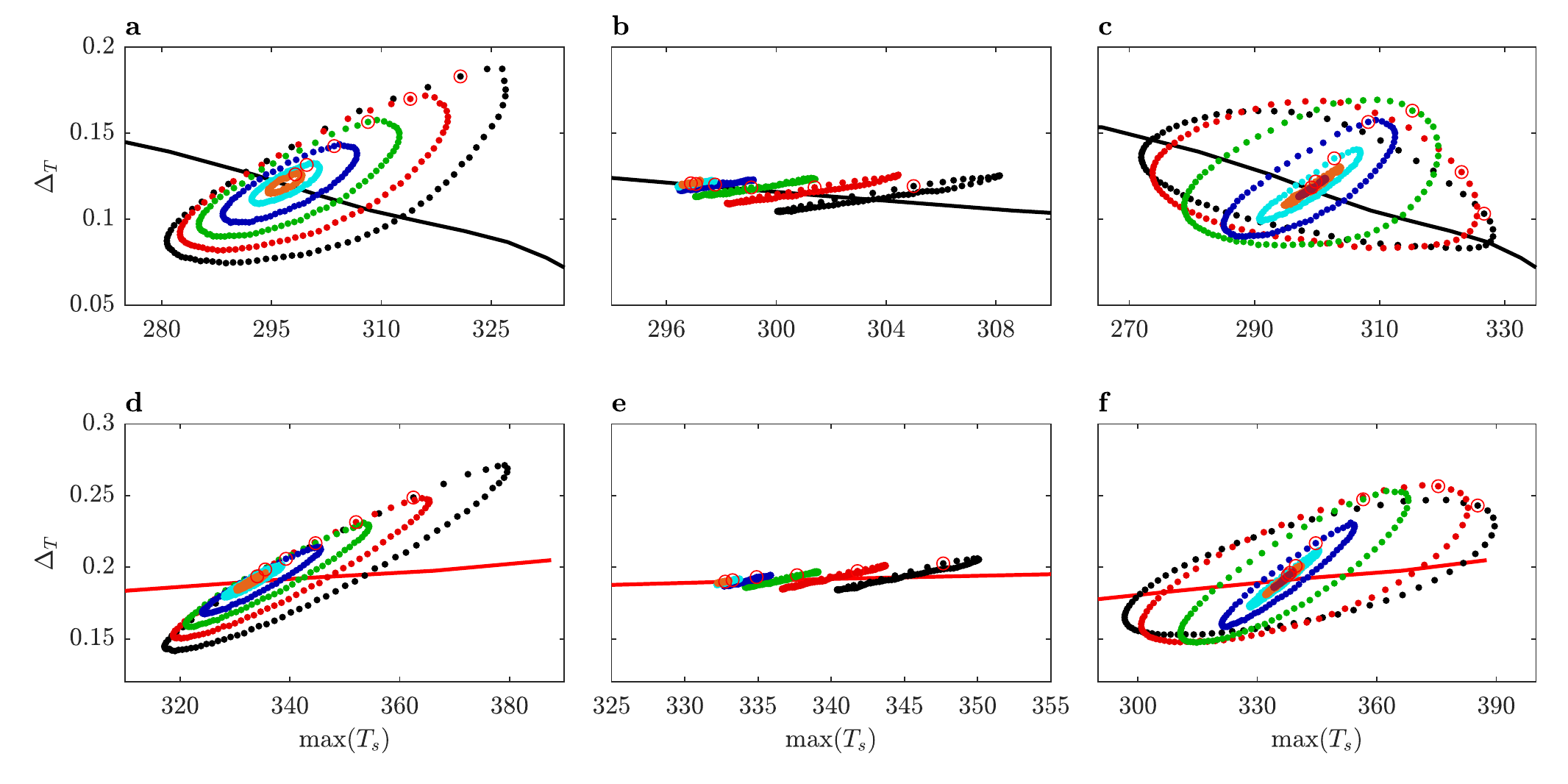}
\caption{$\Delta_T$ as a function of $\max(T_s)$ (K) for different values of eccentricity (0.05 (orange), 0.1 (cyan), 0.2 (blue), 0.3 (green), 0.4 (red), 0.5 (black)), with an Earth-like orbital period (a,d) and an $1/8$ of Earth's orbital period (b,e) and for different orbital periods (0.125 (brown), 0.25 (orange), 0.5 (cyan), 1 (blue), 2 (green), 4 (red), 6 (black) times Earth's orbital period) with $\epsilon=0.3$ (c,f) for moist (first row) and dry (second row) simulations. Red circles represent the first day of the year and the day with maximum insolation (the seasonal cycle moves clockwise). The black and red lines are the line from the perpetual equinox simulations (Fig.~\ref{fig:solar1}a), for moist and dry, respectively. \label{fig:ecc_seas}}
\end{figure*}

\subsection{Temperature response}

Planets with zero obliquity exhibit hemispherical symmetry; due to this symmetry, it is convenient to quantify the surface temperature seasonal cycle using $\Delta_T$ and $\max(T_s)$. In this ($\Delta_T, \ \max(T_s)$) space, the seasonal cycle has a shape of an ellipse, with this shape changing its characteristics depending on the eccentricity and orbital period values, as shown in Figure \ref{fig:ecc_seas}, where the red circle denotes the first day of the year, and the seasonal cycle moves clockwise. Increasing the eccentricity in an Earth-like orbital period (Fig.~\ref{fig:ecc_seas}a,d) results in a stronger seasonal cycle, with most of the response occurring in the cooling period in the seasonal cycle. This is a consequence of the differences between the timescale of the cooling and warming periods during the insolation seasonal cycle; whereas the maximum (minimum) insolation increases (decreases) strongly (weakly) with eccentricity, the time period of this strong warming (weak cooling) becomes shorter (longer) with eccentricity (Fig.~\ref{fig:inso1}a-c), giving the atmosphere less (more) time to adjust to these radiative changes. In comparison, varying the eccentricity for a planet with a short orbital period results in a weak seasonal cycle with almost no differences between the cooling and warming periods (Fig.~\ref{fig:ecc_seas}b,e), an outcome of the radiative changes acting on a faster scale relative to the radiative timescale. In addition to the seasonal cycle changes, there is general warming with eccentricity; this warming is the response to the annual mean insolation. The annual mean insolation increases with eccentricity \citep{bolmont_habitability_2016}, for this reason, if the orbital period is short enough, the climate is forced effectively by the annual mean insolation, resulting in a general warming trend, with the $\Delta_T$ response following the perpetual response (lines in Fig.~\ref{fig:ecc_seas}).

\begin{figure}[ht!]
\centering
\includegraphics[width=\columnwidth]{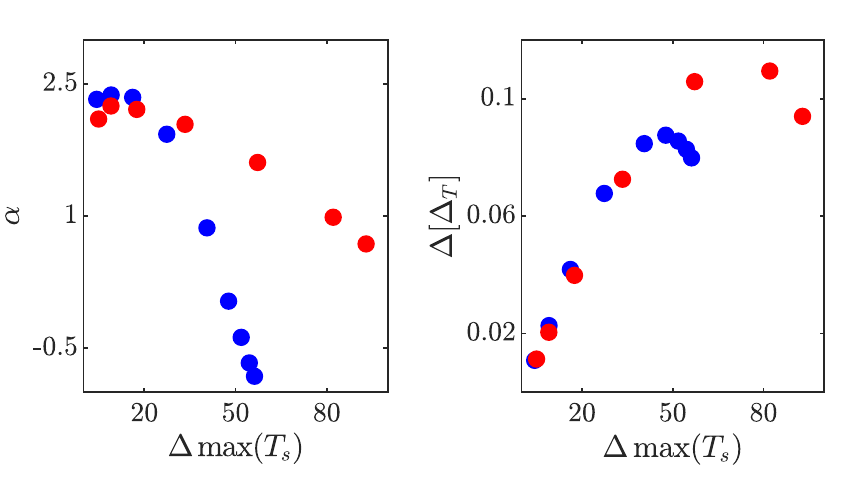}
\caption{$\alpha$ (left panel), the mean slope of $\Delta_T$ with respect to $\max(T_s)$ from Figure \ref{fig:ecc_seas}c,f, and $\Delta[\Delta_T]$ (right panel) the seasonal amplitude of $\Delta_T$ as a function of $\Delta(\max(T_s))$, where an increase in $\Delta(\max(T_s))$ is equivalent to an increase in the orbital period, for moist (blue) and dry (red) cases. Note that for the moist case there are two additional simulations that are not shown in Figure~\ref{fig:ecc_seas} ($\omega=3$ and $5$). \label{fig:angle}}
\end{figure}

The response to changing the orbital period in constant eccentricity (Fig.~\ref{fig:ecc_seas}c,f) has qualitative differences. First, increasing the orbital period alters both the cooling and warming period response, a result of the fact that increasing the orbital period gives the atmosphere more time to adjust in both periods. The second main response of the seasonal cycle to variations in the orbital period is that the general slope of $\Delta_T$ with $\max(T_s)$ changes with the orbital period. We can use the following matrices to quantify this result 
\begin{eqnarray}
    &\Delta\max(T_s)=\max(\max(T_s))-\min(\max(T_s)),\\
    &\Delta[\Delta_T]=\max(\Delta_T)-\min(\Delta_T),\\
    &\alpha=\frac{\Delta_T(\max(\max(T_s)))-\Delta_T(\min(\max(T_s)))}{\Delta\max(T_s)},
\end{eqnarray}
where $\Delta\max(T_s)$ and $\Delta[\Delta_T]$ represent the seasonal amplitude of changes in $\max(T_s)$ and $\Delta_T$, respectively. Note that an increase in $\Delta\max(T_s)$ is equivalent to an increase in the orbital period. $\alpha$ represents the mean slope of $\Delta_T$ as a function of $\max(T_s)$. Both $\Delta[\Delta_T]$ and $\alpha$ are non-monotonic with the orbital period, where both exhibit an increase in short orbital periods and a decrease in longer ones (Fig.~\ref{fig:angle}).

A good starting point for understanding the non-monotonic dependence of $\alpha$ and $\Delta[\Delta_T]$ on the orbital period is to consider the extremes. The first extreme is a very short orbital period; in this case, the radiative timescale is much longer than the timescale at which the radiative changes occur, resulting in a climate that is effectively forced by the annual mean forcing and can be represented as a point in Figure~\ref{fig:ecc_seas}. In the second extreme, the 'infinite orbital period', the radiative changes occur over long timescales, allowing the atmosphere to relax to the seasonal forcing. This is equivalent to changing the solar constant in a perpetual equinox, and as a result the seasonal cycle will coincide with the line representing the perpetual case (black and red lines in Fig.~\ref{fig:ecc_seas}). Using these extreme cases, it becomes clear why the seasonal cycle in long orbital periods approaches the perpetual line, as does the decrease in $\alpha$ and $\Delta[\Delta_T]$ at long orbital periods. The remaining question is, What controls the shape of the seasonal cycle in short-to-moderate orbital periods?

\subsubsection{Simple energy balance model}

\begin{figure}[ht!]
\centering
\includegraphics[width=\columnwidth]{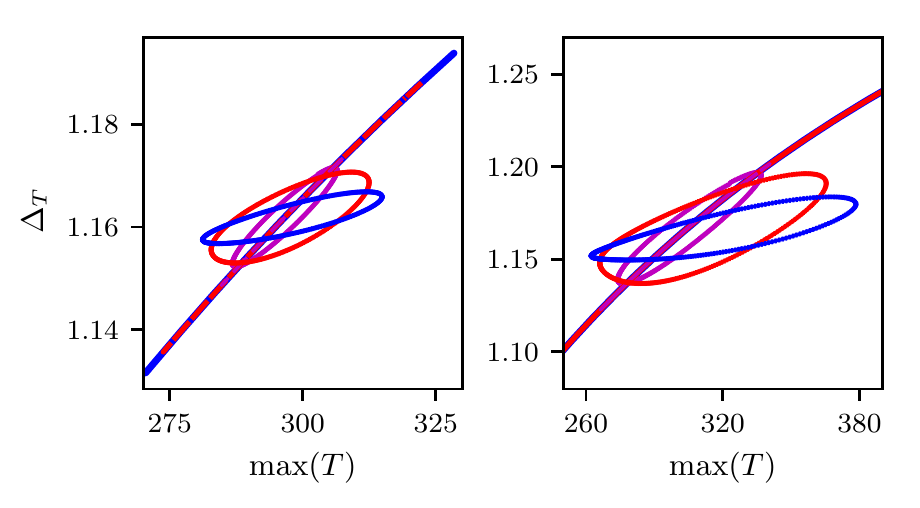}
\caption{Solution of the simple EBM (Eq.~\ref{eq:EBM_sol}) for $\varepsilon=0.1$ (left panel) and $0.3$ (right panel) with different orbital period values (1/8 (purple), 1 (red), 4 (blue)). The lines are solutions with $\tau_{\rm{rad}}$ taken to be constant with latitude, with a value of $30$ days, the ellipses are solutions for a latitude-dependent $\tau_{\rm{rad}}$, given by Equation \ref{eq:rad_time}. \label{fig:ebm}}
\end{figure}

In order to understand what controls the temperature seasonal cycle in short-to-moderate orbital periods we construct a simple energy balance model. As mentioned, for long orbital periods, $\alpha$ starts to decrease as a result of the atmosphere having enough time to respond to the radiative changes, and the dynamics to change the temperature structure. Given this, we can assume that in short orbital periods, the dynamics do not have enough time to alter the seasonal cycle temperature structure significantly, and the main process is a radiative one. Based on this argument, consider a simple dry, non-diffusive energy balance model,
\begin{equation}\label{eq:EBM1}
    C\frac{dT}{dt}=Q-\sigma T^4,
\end{equation}
where $C$ is the heat capacity, and $Q$ is the insolation, with both $Q$ and $T$ are a function of time and latitude. To simplify this even further, similar to the derivation in \citet{Mitchell_2014}, we can assume that the annual mean is relaxed to the annual mean forcing. This assumption is a direct result of taking the annual mean of Equation \ref{eq:EBM1}, and can be justified from Figure \ref{fig:ecc_seas}, where the center of each ellipse falls on, or is close to, the perpetual equinox line, suggesting that the annual mean is relaxed to the relevant perpetual equinox scenario. Using this assumption, the temperature and insolation can be decomposed into
\begin{eqnarray}
    T=\overline{T}+T',\\
    Q=\overline{Q}+Q',
\end{eqnarray}
where
\begin{equation}
    \overline{T}=\left(\frac{\overline{Q}}{\sigma}\right)^{1/4}.
\end{equation}
Here, the bar and prime denote the time mean and deviations from the time mean respectively. Assuming that $T'\ll \overline{T}$, we can linearize Equation \ref{eq:EBM1} giving
\begin{equation}\label{eq:linEBM}
    C\frac{dT'}{dt}=Q'-4\sigma\overline{T}^3T'.
\end{equation}
As described in \citet{Mitchell_2014}, solutions to this equation for $T'$ have both a phase lag and amplitude reduction relative to the forcing term $Q'$. Equation \ref{eq:linEBM} is a linear ordinary differential equation with the general solution
\begin{equation}
    T'=\left[\int \frac{Q'}{C}\exp\left(\frac{t}{\tau_{\rm{rad}}}\right)dt+T_0\right]\exp\left(-\frac{t}{\tau_{\rm{rad}}}\right),
\end{equation}
where $\tau_{\rm{rad}}$ is the radiative timescale (as in Eq.~\ref{eq:rad_time}, substituting $T_e$ with $\overline{T}$) and $T_0$ is the initial condition\footnote{$T_0$ is given by calculating $T'$ with a random value for $T_0$ for one year and using the last step from this calculation as the initial condition for the solution shown in Figure \ref{fig:ebm}.}. In order to illustrate the role of the orbital period, we can write $t \rightarrow \omega t'$, where $\omega$ is the orbital period; using this notation, we can write the temperature solution 
\begin{eqnarray}\label{eq:EBM_sol}
    &T=\left(\frac{\overline{Q}}{\sigma}\right)^{1/4} + \nonumber\\ &\left[\int \frac{\omega Q'}{C}\exp\left(\frac{\omega}{\tau_{\rm{rad}}}t'\right)dt'+T_0\right]\exp\left(-\frac{\omega}{\tau_{\rm{rad}}}t'\right).
\end{eqnarray}
The nature of the solution strongly depends on the latitudinal structure of $\tau_{\rm{rad}}$. If $\tau_{\rm{rad}}$ is taken to be the same at all latitudes, the solution is simply a straight line, that becomes longer with the orbital period (Fig.~\ref{fig:ebm}). However, taking $\tau_{\rm{rad}}$ with a latitudinal structure (as in Equation $\ref{eq:rad_time}$, with $\overline{T}$), the temperature solution becomes qualitatively similar to the GCM solution. This result suggests that, at least for the short and moderate orbital periods where $\omega/\tau_{\rm{rad}}$ is small enough, the eccentricity seasonal cycle can be explained using these radiation balance arguments. Note that this statement is true only for the seasonal cycle response, that is considered a perturbation around the mean state that is strongly affected by dynamics and other processes. Once $\omega/\tau_{\rm{rad}}$ is large enough, the atmosphere has more time to respond to the radiative changes, and other dynamical and nonlinear radiative effects come into play. Additionally, the simple solution's (Eq.~\ref{eq:EBM_sol}) dependence on the latitudinal structure of $\tau_{\rm{rad}}$ underlines the importance of the latitudinal structure of the radiative timescale for the response of the atmosphere to the eccentricity seasonal cycle.

\subsection{Circulation response}
Merdional temperature gradients affect the atmospheric general circulation. The balance between the meridional temperature gradients and the zonal mean shear, for rapidly rotating planets, is given by the thermal wind balance 
\begin{equation}\label{eq:tw_balance}
    f\frac{\partial u}{\partial p}=\frac{R_d}{p}\frac{1}{a}\left(\frac{\partial T}{\partial \phi}\right)_p,
\end{equation}
where $f=2\Omega\sin\phi$ is the Coriolis parameter, with $\Omega$ the rotation rate. The subscript $p$ in the last term in Equation \ref{eq:tw_balance} denotes that the derivative is taken over isobaric surfaces. The thermal wind balance is the first order balance for an atmosphere in hydrostatic balance on fast rotating planets \citep{vallis_2017, galanti_kaspi_tziperman_2017} implying that where steeper meridional temperature gradients are balanced by stronger vertical zonal wind shear. In addition to the effect on the zonal winds, the mean meridional circulation is also strongly affected by the meridional temperature gradients, where the Hadley circulation becomes stronger and wider as the meridional temperature gradient increases \citep{held1980}. 

\begin{figure*}[ht!]
\centering
\includegraphics[width=0.8\paperwidth]{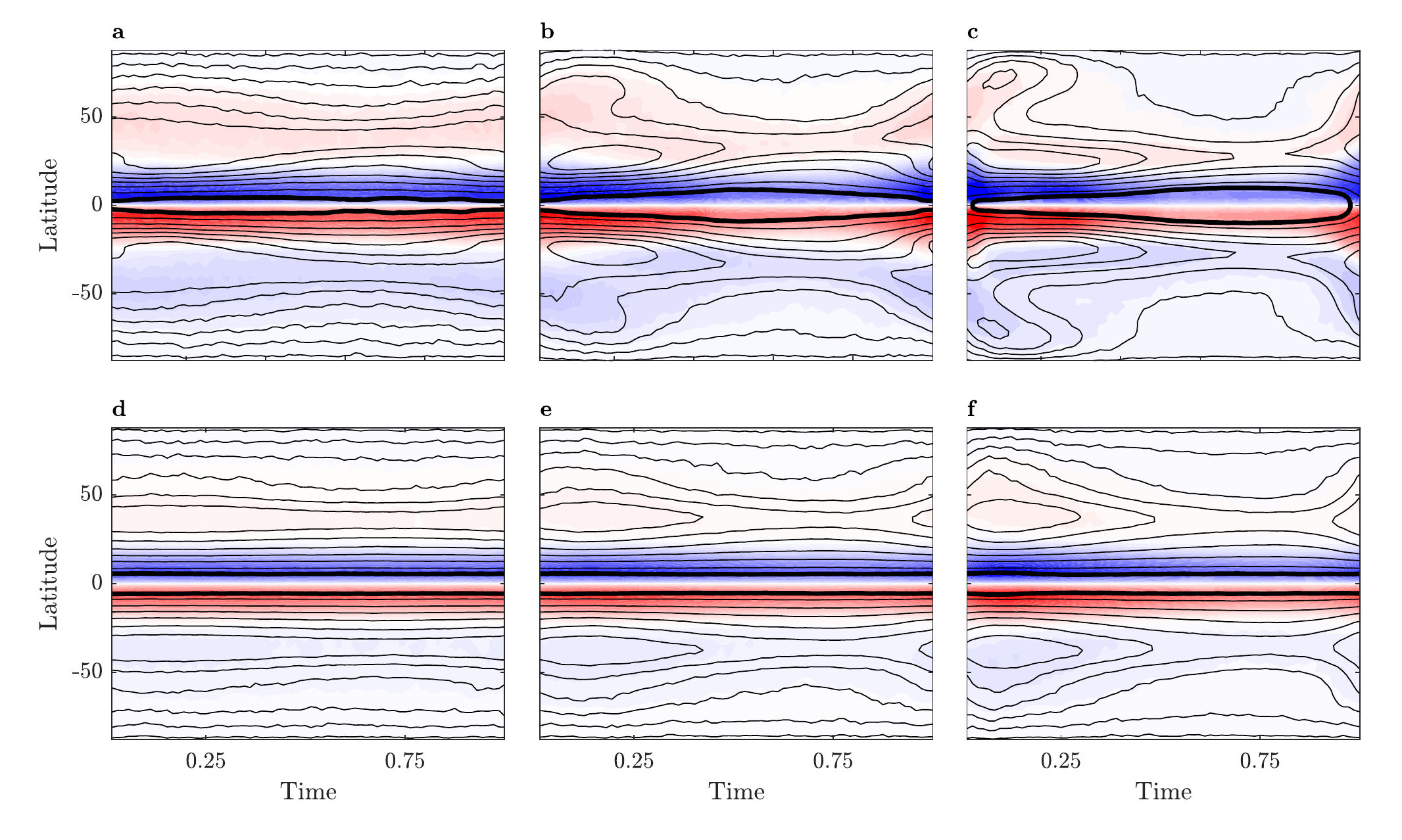}
\caption{Comparison of the moist (top row) and dry (bottom row) seasonal cycle of the meridional circulation, averaged vertically between $400-600$ hPa (shading, blue means northward flow in the upper branch of the circulation), and the zonal mean zonal wind vertically averaged between $100-500$ hPa (contours) for eccentricities: $0.1$ (a,d) $0.3$ (b,e) $0.5$ (c,f). \label{fig:stream_ecc}}
\end{figure*}

Both the meridional streamfunction and the zonal wind exhibit a seasonal cycle that is more pronounced in the moist case (Fig.~\ref{fig:stream_ecc}), consistent with the perpetual case where the dynamics have a more complex dependence on $S_0$ (Fig.~\ref{fig:solar1}). However, in contrast to the perpetual moist case, where, for example, warmer climate results in a weaker circulation, this is not the case for the seasonal cycle, a result of the different dependence of $\Delta_T$ on $\max(T_s)$. Also, the only case where equatorial superrotation persist for the seasonal cycle, is for $\varepsilon=0.5$ (Fig.~\ref{fig:stream_ecc}c); however, it is correlated with high $\Delta_T$, unlike the perpetual case, suggesting that a different mechanism is responsible for the transition to superrotation in the seasonal cycle case.

In rotating atmospheres, two general processes can accelerate a westerly (prograde) jet stream, and both involve a source of angular momentum for the prograde flow. The first mechanism relates to the poleward transfer of air from the warm tropics to higher latitudes (e.g., the Hadley circulation). If, in this process, the poleward traveling air conserves its angular momentum, starting with a zero zonal mean zonal wind at the equator, the angular momentum conserving wind \citep{vallis_2017} is
\begin{equation}
    u_m=\Omega a\frac{\sin^2\phi}{\cos\phi}.
\end{equation}
This process will result in a prograde jet at the edge of the Hadley circulation. This type of jet is called a thermally driven jet, and on Earth, also the subtropical jet. Note that this is an ideal description of this process, whereas in reality, turbulence and other processes that are neglected in this ideal scheme could be relevant \citep[e.g.,][]{singh2016, singh2017}.

The second process that can contribute to the acceleration of prograde jets relates to wave braking in the atmosphere. In the midlatitudes, where the temperature gradients are concentrated, baroclinic\footnote{Baroclinicity is the measure of the misalignment of density and pressure surfaces; when these surfaces align, the fluid is considered barotropic.} instability develops, creating disturbances in this region of the atmosphere. It can be shown, using potential vorticity (PV) and angular momentum conservation arguments, that disturbances in this region converge momentum in the disturbance latitudes, resulting in a prograde jet \citep{vallis_2017}. This type of jet is called an eddy-driven jet. 

On Earth, these two processes occur in proximity to each other, resulting mainly in a merged thermally- and eddy-driven jet. However, the jet characteristics change during the seasonal cycle \citep{lachmi2014, vallis_2017, yuval2018}. Looking at other planets, mainly the gas giants, Saturn and Jupiter, have multiple jets in each hemisphere \citep[e.g.,][]{ingersoll1990}. Generally, the number of jets in each hemisphere for a given planetary atmosphere relates to the typical eddy size and inverse energy cascade length scale \citep{rhines1975,rhines1979,chemke2015a,chemke2015b}. More specifically, the inverse energy cascade scale, i.e., the Rhines scale, $L_R$, is defined as
\begin{equation}
    L_R=\left(\frac{2U}{\beta}\right)^{1/2},
\end{equation}
where $U$ is a measure of the zonal wind (the root mean square velocity, often taken as the rms of eddy velocities \citep{rhines1975}) and $\beta=2\Omega\cos\phi/a$, is the meridional derivative of the Coriolis parameter. An estimate for the number of jets is given by \citep{wang2018}
\begin{equation}
    N_j\approx\frac{a}{4L_R}.
\end{equation}
\citet{wang2018} and \citet{lee2005} used different forms to estimate $L_R$, finding that $N_j\propto(\Delta\theta)^{-1/2}$, where $\Delta\theta$ is the equator-to-pole potential temperature difference. \citet{wang2018} also tested other estimates for $L_R$ and showed that it yields a similar result.

\begin{figure*}[ht!]
\centering
\includegraphics[width=0.8\paperwidth]{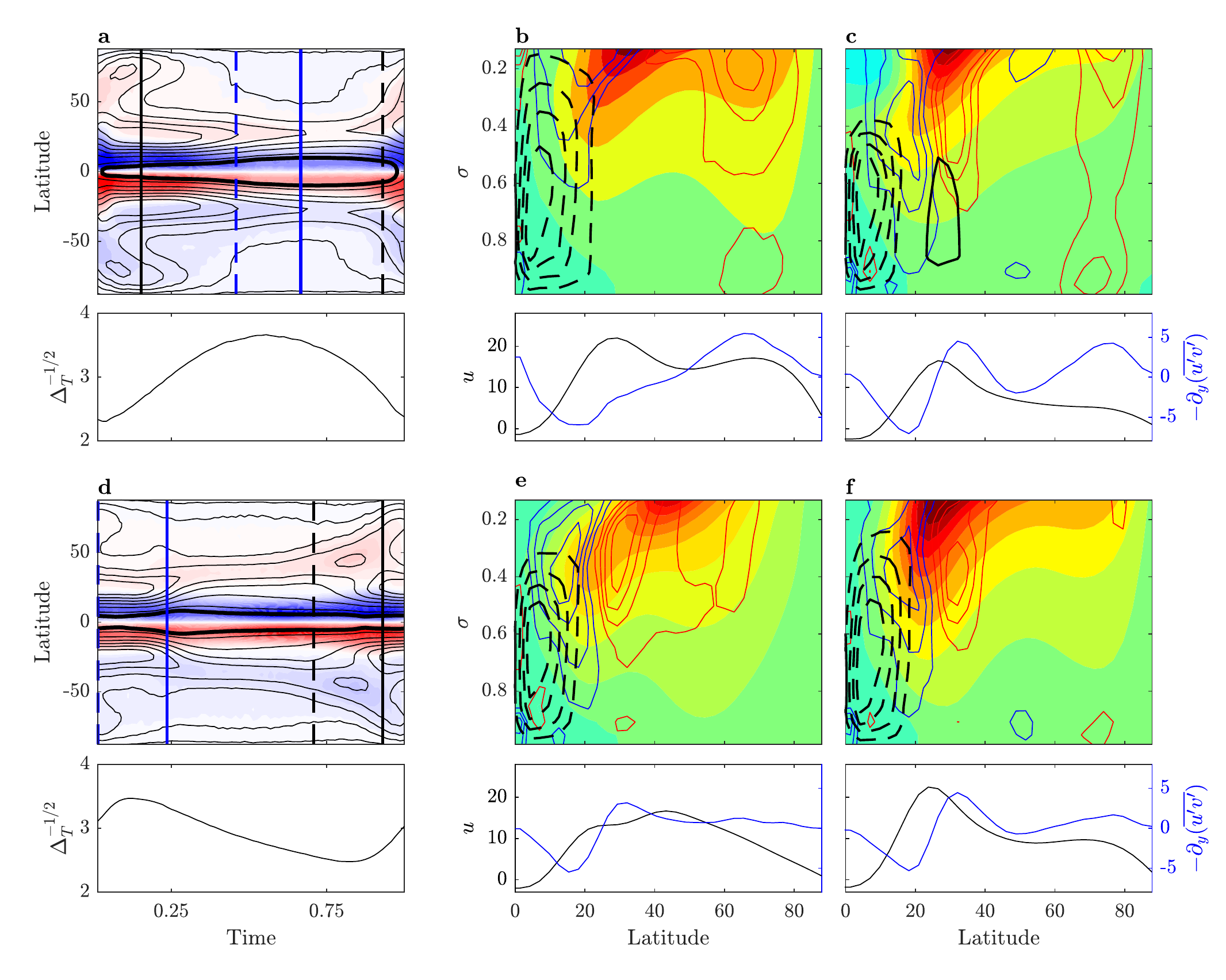}
\caption{The seasonal cycle of the zonal mean zonal wind ($\overline{u}$) and mean meridional circulation ($\psi$). Panels a and d show the seasonal cycle of $\overline{u}$ and $\psi$ (top, similar to Fig.~\ref{fig:stream_ecc}) and $\Delta_T^{-1/2}$ (bottom) for the $\epsilon=0.5$ and $\omega=1$ case, and the $\epsilon=0.3$ and $\omega=4$ case, respectively. Vertical lines represent the time average for panels b and e (black), and c and f (blue), with the dashed line representing the beginning of the average period. Panels b, c, e, and f: top shows the time mean of the zonal mean zonal wind (shading), mean meridional circulation (black contours, dashed lines are clockwise circulation) and zonal mean eddy momentum flux convergence ($-\partial_y(\overline{u'v'})$, red and blue contours, red is for convergence); bottom shows the vertically averaged $\overline{u}$ (black, $\rm{m \ s^{-1}}$) and $-\partial_y(\overline{u'v'})$ (blue, $10^{-6}\rm{m \ s^{-2}}$). \label{fig:u_ecc}}
\end{figure*}

As the circulation seasonal cycle response seems to be relatively weak (Fig.~\ref{fig:stream_ecc}), it is useful to look at the extreme cases. Given the qualitatively different response of the temperature to changes in the eccentricity and the orbital period, we compare the circulation response between two simulations, the first is a $\varepsilon=0.5$ and $\omega=1$ simulation (hereafter referred to as high eccentricity simulation), and the second is a $\varepsilon=0.3$ and $\omega=4$ simulation (hereafter referred to as long orbital period simulation).

The mean meridional circulation gets stronger as $\Delta_T$ increases in both cases (Fig.~\ref{fig:u_ecc}a,d), and this relation can be explained using axisymmetric arguments \citep[as mentioned earlier in this manuscript;][]{held1980}. The circulation deepens as the surface temperature rises (black contours in Fig.~\ref{fig:u_ecc}b,c,e,f top panels), which is a similar response to that of the tropopause height in the perpetual case (Fig.~\ref{fig:solar1}d). Note that qualitative differences in the temperature response result in a qualitative difference in the circulation structure. More specifically, in the high eccentricity simulation, a high surface temperature is accompanied by a large $\Delta_T$ (Fig.~\ref{fig:ecc_seas}a), resulting in deeper and stronger circulation at large $\Delta_T$ (Fig,~\ref{fig:u_ecc}b-c). However, in the long orbital period simulation, a large $\Delta_T$ comes along with a relatively low surface temperature (Fig.~\ref{fig:ecc_seas}c), resulting in a more complex response, where the higher circulation is weaker (Fig.~\ref{fig:u_ecc}e-f). 

The characteristics of the zonal mean zonal wind, $\overline{u}$, also change during the seasonal cycle. As mentioned, there are two types of jets, the thermally driven jet, associated with the Hadley circulation, and the eddy-driven jet; on Earth, they are generally merged. There are several ways to distinguish between the two. First, the thermally driven jet is located at the edge of the Hadley circulation, whereas the eddy-driven one is associated with eddy momentum flux convergence. Second, their vertical structure is different, where the thermally driven jet has a more baroclinic structure, and the eddy-driven jet, a more barotropic structure \citep{vallis_2017}.

During the seasonal cycle of both simulations, there is a transition from one to two jets (Fig.~\ref{fig:u_ecc}). Following the $(\Delta\theta)^{-1/2}$ scaling, using $\Delta_T^{-1/2}$ as a proxy for it, $\Delta_T^{-1/2}$ correlates well with the number of eddy-driven jets. For both cases, in minimum values of $\Delta_T^{-1/2}$, there is only one eddy-driven jet (Fig.~\ref{fig:u_ecc}b,e); however, in the high eccentricity case, the eddy driven and thermally driven jets are separated (Fig.~\ref{fig:u_ecc}b). In contrast, in the long orbital period case, there is only one merged jet (Fig.~\ref{fig:u_ecc}e). Around the maximum values of $\Delta_T^{-1/2}$, in both simulations, there are two eddy driven jets, one merged with the thermally driven jet and the other (relatively weak) at higher latitudes (Fig.~\ref{fig:u_ecc}c,f). Another difference between the two simulations is that for the high eccentricity simulation there is a short period with equatorial superrotaiton (Fig.~\ref{fig:u_ecc}a), which does not happen in the long orbital period simulation. This superrotation happens when $\Delta_T^{-1/2}$ reaches its minimum values, i.e., high $\Delta_T$ values. A manifestation of this superrotation can be seen in Figure \ref{fig:u_ecc}b, where there is a weak eddy momentum flux convergence at the equator. Another distinctive feature, in this case, is that the midlatitude eddy momentum flux convergence is more poleward than all other cases (Fig.~\ref{fig:u_ecc}). This poleward shift of the eddy momentum flux convergence may suggest that the Rossby waves responsible for the acceleration of this jet transport momentum from the subtropics, instead of the tropics, allowing momentum to converge at the equator. This is similar to the mechanism suggested by \citet{mitchell2010} for high and intermediate thermal Rossby numbers,
\begin{equation}\label{eq:thermal_rossby}
    R_o=\frac{2gH\Delta_T}{\Omega^2 a^2},
\end{equation}
which can be relevant in this case, as $\Delta_T$ is large. However, as already noted, the details of the superrotaion mechanism are beyond the scope of this study.

\section{The seasonal cycle on a planet in an eccentric orbit with non-zero obliquity} \label{sec:obliq_ecc}
\begin{figure*}[htb!]
\centering
\includegraphics[width=0.8\paperwidth]{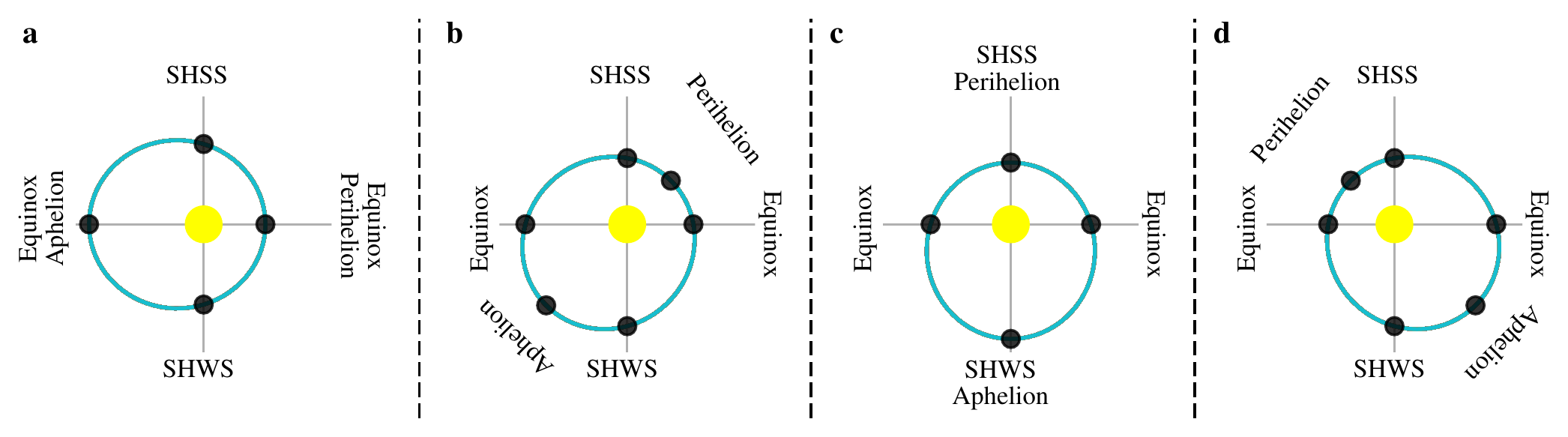}
\caption{Distinctive orbital configurations for an orbit with non-zero obliquity and eccentricity. (a) Perihelion and equinox are aligned ($\Pi=0^{\circ}$). (b) Perihelion after equinox and before southern hemisphere summer solstice (SHSS, $\Pi=45^{\circ}$). (c) Perihelion and SHSS are aligned ($\Pi=180^{\circ}$). (d) Perihelion after southern hemisphere winter solstice (SHWS) and before equinox ($\Pi=135^{\circ}$). \label{fig:orbit}}
\end{figure*}
Planets with non-zero obliquity (tilted planets) experience a seasonal cycle of the insolation meridional structure; during this seasonal cycle, the maximum insolation shifts from one hemisphere to the other. In addition, the maximum insolation increases and becomes more poleward with obliquity. The obliquity seasonal cycle can be characterized by two periods during the orbital cycle, equinox and solstice. Equinox is when the maximum insolation is at the equator, which occurs twice in a cycle, in this study, at stellar longitudes $0^{\circ}$ and $180^{\circ}$. Solstice is when the insolation peaks at the most poleward latitude, once in each hemisphere during the seasonal cycle; in this study, as a matter of convention, the southern hemisphere summer solstice (SHSS) is at stellar longitude $90^{\circ}$, and the summer hemisphere winter solstice (SHWS, alternatively the northern hemisphere summer solstice) is at stellar longitude $270^{\circ}$. 

\begin{figure*}[htb!]
\centering
\includegraphics[width=0.8\paperwidth]{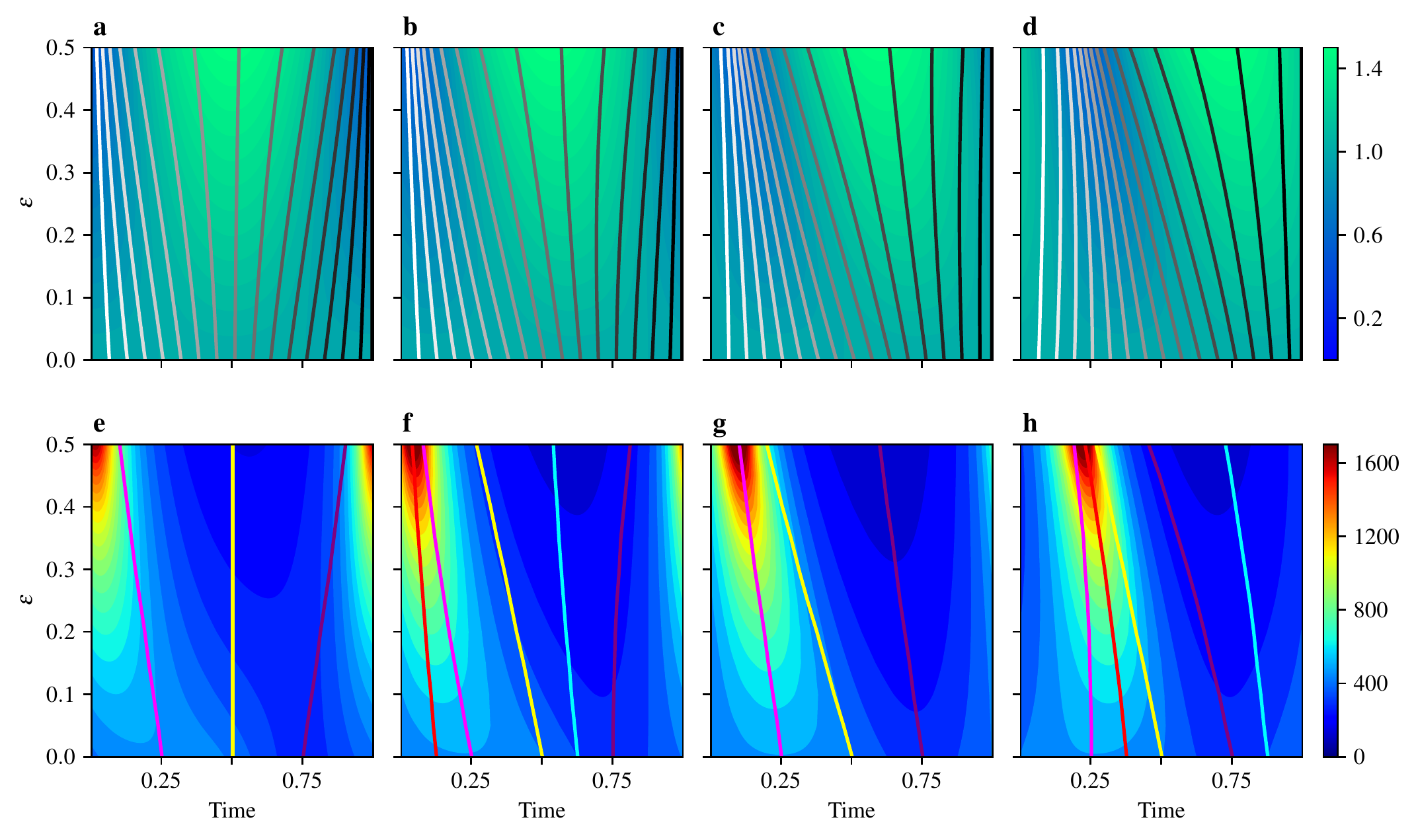}
\caption{Top row: The normalized distance from the host star (shading) and the angle relative to equinox (contours, color values are from $0^{\circ}$ (white) to $359^{\circ}$ (black)) as a function of eccentricity and time, for the four configurations shown in Figure \ref{fig:orbit}, respectively. Bottom row: The insolation at latitude $23^{\circ}$ in the southern hemisphere (shading, W m$^{-2}$); magenta line is for SHSS, yellow line for the second equinox ($180^{\circ}$), purple line for SHWS, red line for perihelion and cyan line for aphelion. All plots are with an obliquity of $23^{\circ}$.  \label{fig:forcing}}
\end{figure*}

For a tilted planet in an elliptical orbit (non-zero eccentricity), the stellar longitude of perihelion ($\Pi$), e.g., the position where the planet is closest to its host star, relative to equinox is important. Note that in this study, the stellar longitude of perihelion also denotes its phase with equinox (as equinox remains at stellar longitude $0^{\circ}$). Due to the importance of the perihelion position, it is essential to distinguish between different orbital configurations, which can be generally classified into four types: Alignment of perihelion with equinox ($\Pi=0^{\circ}$, Fig.~\ref{fig:orbit}a) perihelion is after equinox and before the SHSS ($\Pi=45^{\circ}$, Fig.~\ref{fig:orbit}b), perihelion aligned with SHSS ($\Pi=90^{\circ}$, Fig.~\ref{fig:orbit}c) and perihelion after SHSS and before equinox ($\Pi=135^{\circ}$, Fig.~\ref{fig:orbit}d). Note that perihelion values $180^{\circ}-315^{\circ}$ are a mirror image on the other hemisphere (assuming hemispheric symmetry). For given obliquity and eccentricity values, different perihelion positions will result in different insolation seasonal cycles depending on the different orbital parameters (Fig.~\ref{fig:inso1}g-i, \ref{fig:forcing}e-h). 

In an eccentric orbit, the orbital velocity is not constant during the orbital period and depends on the planet's distance from the star, with the orbital velocity increasing as the planet comes closer to its host star \citep{lissauer_2013}. As eccentricity increases, this effect is magnified, and equal orbital distances will pass in a different timescales. This effect is illustrated in the first row of Figure \ref{fig:forcing}, where, as the distance to the star becomes shorter (blue colors), the contours, representing the orbital angle, become denser. This means that changes in the insolation will occur over different timescales during the orbital period, depending on the orbital configuration (Figs.~\ref{fig:inso1},\ref{fig:forcing}e-h), where, in general, as eccentricity increases, stronger forcing will occur over shorter periods (Fig.~\ref{fig:forcing}e-h). 

To summarize, the combination of eccentricity and obliquity introduces a new degree of freedom, which is the relative position of the perihelion with respect to equinox. The complexity is emphasized by insolation changes occurring over different timescales, depending on the orbital configuration (Fig.~\ref{fig:forcing}). As a result, the study of the climate on a tilted planet in an eccentric orbit depends on a wide range of parameters. The purpose of this section is to show preliminary results of the climate dependence on obliquity, eccentricity, and perihelion, which should serve as a baseline for future studies, give some constraints on the atmospheric circulation, and discuss the importance and relevance of the different planetary parameters that can affect the climate response to changes in the orbital configuration.

\subsection{Temperature response}
\begin{figure*}[htb!]
\centering
\includegraphics[width=0.8\paperwidth]{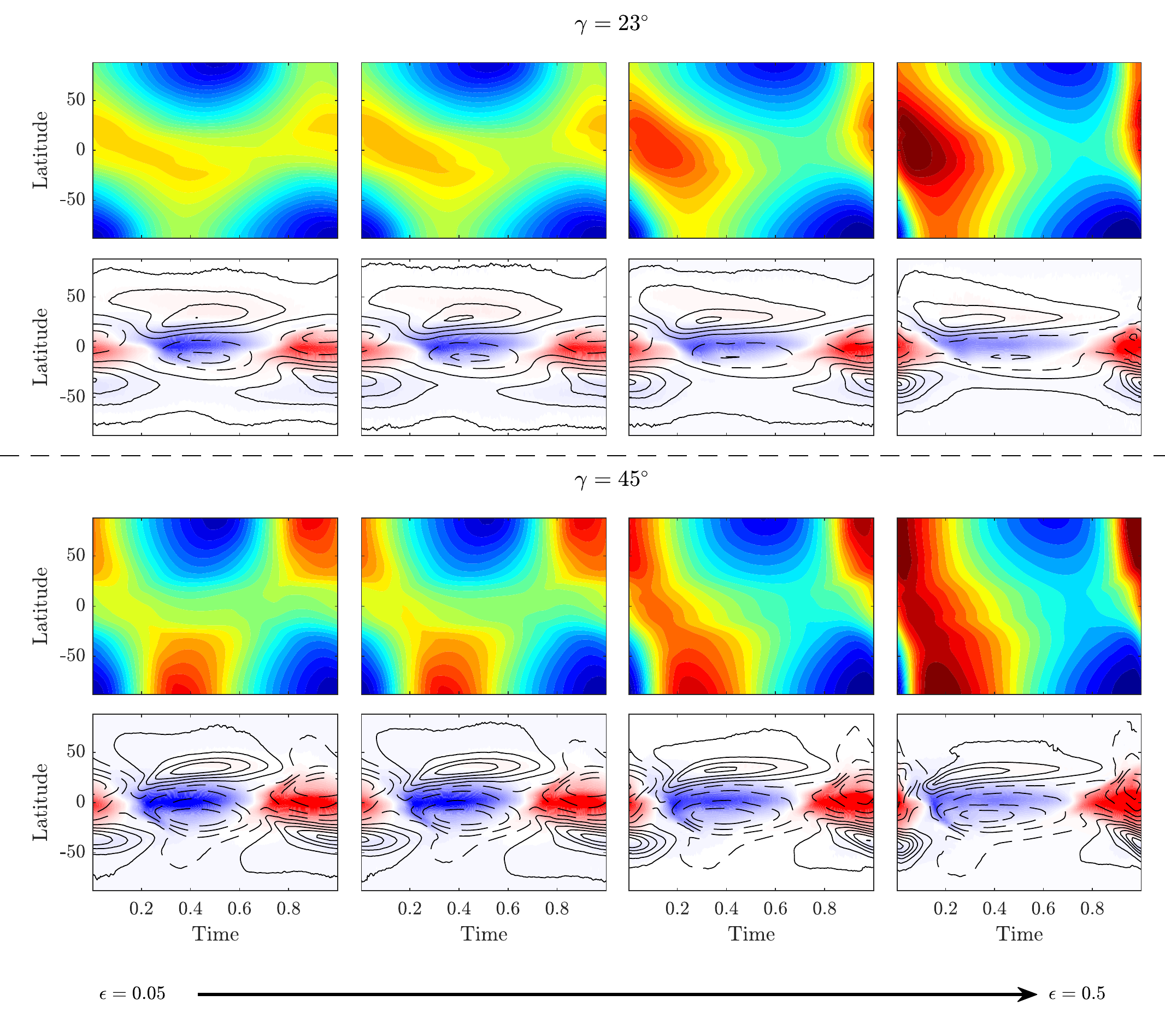}
\caption{The seasonal cycle dependence on eccentricity for an obliquity of $23^{\circ}$ (top half) and $45^{\circ}$ (bottom half). Top row shows the seasonal cycle of the surface temperature, with colors ranging from $250^{\circ}$ to $320^{\circ}$ K. The bottom row shows the seasonal cycle of the mean merional circulation, vertically averaged between $400$ and $600$ hPa (shading, blue means northward flow in the upper branch of the circulation), and the zonal mean zonal wind, vertically averaged between $100$ and $500$ hPa. Eccentricity increases from left to right, with values of $0.05, \ 0.1, \ 0.3, \ 0.5$, and $\Pi=0^{\circ}$ in all panels. \label{fig:ecc_ob}}
\end{figure*}

The insolation of a tilted planet with an eccentric orbit is a function of three parameters: Obliquity ($\gamma$), eccentricity ($\varepsilon$), and perihelion ($\Pi$). This dependence implies that determining the climate on such a planet is a complex problem that depends on a large number of parameters. In order to examine the role of the different parameters, we present a series of simulations in which we vary these three parameters. Of these three parameters, the most studied in the context of the atmospheric circulation is obliquity \citep[e.g.,][]{guendelman2019, ohno_atmospheres_2019, lobo2019}. These studies show the influence of the seasonal cycle on the climate and the importance of considering other parameters that relate to the atmospheric radiative timescale response when taking into account seasonal changes \citep{guendelman2019}. Increasing the obliquity results in a stronger seasonal cycle of the insolation (Fig.~\ref{fig:inso1}d-f), which, in turn, yields a strong temperature and circulation seasonal cycle that increases with obliquity \citep{guendelman2019, lobo2019}.

\begin{figure*}[htb!]
\centering
\includegraphics[width=0.8\paperwidth]{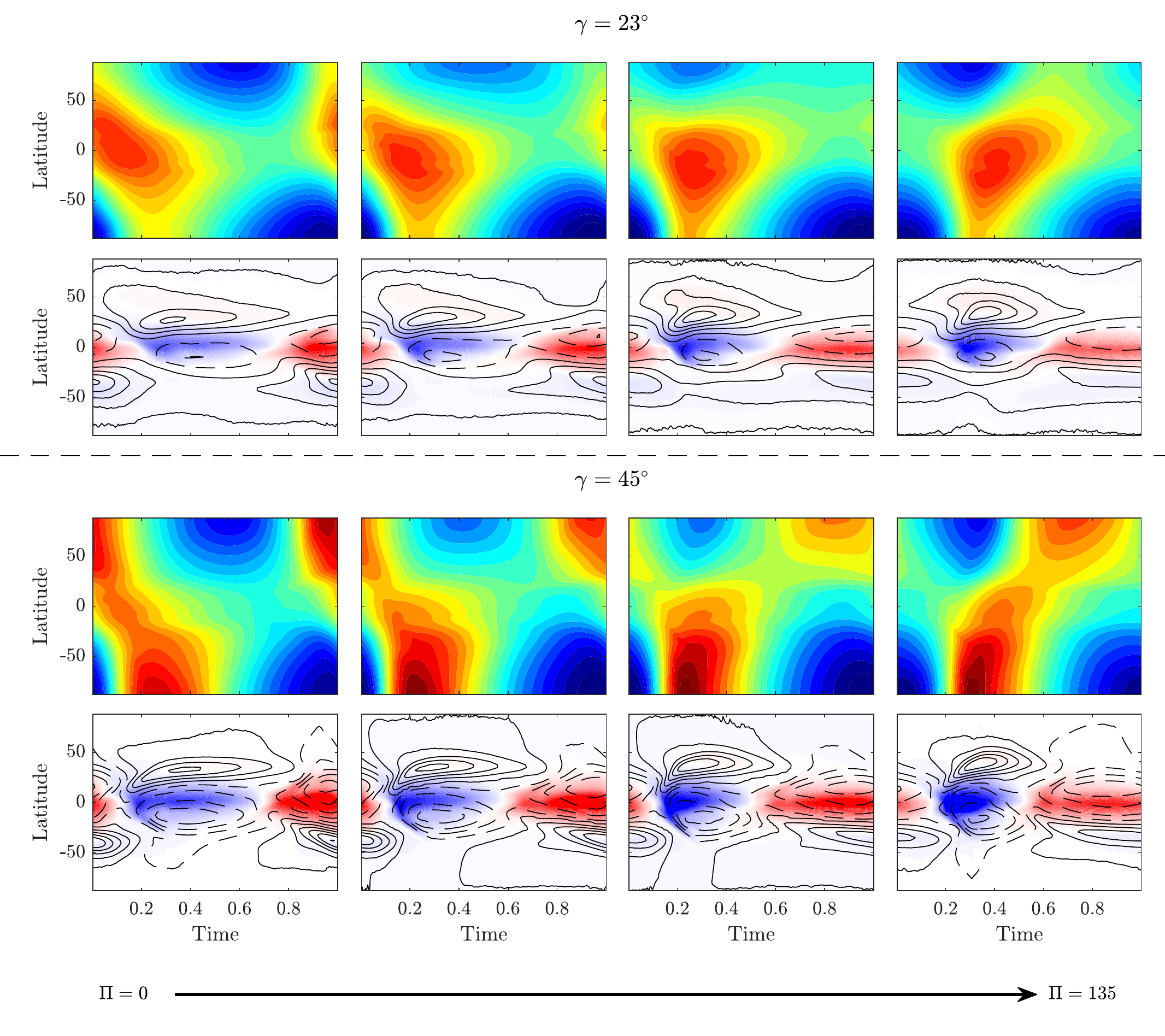}
\caption{Similar to Figure \ref{fig:ecc_ob}, for perihelion dependence. Perihelion increases from left to right, with values of $0, \ 45, \ 90, \ 135$, with $\varepsilon=0.3$ in all panels. \label{fig:ecc_per}}
\end{figure*}

A dominant feature of the surface temperature response is the time delay between the insolation and the temperature response. For example, at equinox (first day in all simulations), the insolation peak at the equator does not coincide with maximum temperature at the equator, which occurs later on. This time lag is due to the atmospheric and surface radiative timescales (a slab ocean with a $10$ m mixed layer). In addition, this time lag is not the same in all the simulations and depends on the eccentricity and perihelion values (Figs.~\ref{fig:ecc_ob} and \ref{fig:ecc_per}). This is a result of the dependence of the time period in which radiative changes occur during the seasonal cycle on the eccentricity and perihelion. In close proximity to perihelion, the radiative changes accelerate with increasing eccentricity. In addition, close to perihelion, there is usually a peak in the insolation (Fig.~\ref{fig:forcing}e-h), which increases with eccentricity and, in response, the atmosphere gets warmer, resulting in a shorter radiative timescale (Eq.~\ref{eq:rad_time}).

As a result of the eccentric orbit, there is an asymmetry between different hemispheres at similar seasons (for example, differences between summer in the northern hemisphere to summer in the southern hemisphere or differences between the two equinoxes). These differences manifest in the seasons' mean temperature, meridional temperature gradient, and the length of each season. For example, for perihelion at equinox, an increase in eccentricity will result in one short and warm equinox, while the other equinox will be long and cold. The short equinox also means a fast transition between one solstice to the other and, as a result of the atmosphere and surface thermal inertia, a difference between the two solstice seasons' duration and strength (Fig.~\ref{fig:ecc_ob}), although the insolation is the same in both (Fig.~\ref{fig:inso1}g). In contrast to the seasonal cycle of a tilted planet in a circular orbit, where the seasonal maximum and minimum temperatures are at approximately the same time \citep[in opposite hemispheres,][]{guendelman2019}, in an eccentric orbit, the seasonal maximum and minimum temperatures are separated in time, with this separation increasing with eccentricity (Fig.~\ref{fig:ecc_ob}). Note that this effect is strongly dependent on the perihelion position, where for perihelion at solstice, there is an alignment in time between the seasonal maximum and minimum temperatures (Fig.~\ref{fig:ecc_per}). In this case, the main effect of eccentricity is the asymmetry between the hemispheres, where one experiences an extreme winter and summer, while in the other hemisphere, the winter and summer are moderate. This perihelion dependence leads to the conclusion that the temperature response strongly depends on both the latitudinal radiative timescale, i.e., the difference in the radiative timescale of different latitudes due to the difference in temperature (Eq.~\ref{eq:rad_time}), and the timescale at which radiative changes occur during the seasonal cycle, which depends on the eccentricity and perihelion.

\subsection{Circulation response}
The zonal mean zonal wind (contours in Figs~\ref{fig:ecc_ob} and \ref{fig:ecc_per}) response follows the temperature response (Eq.~\ref{eq:tw_balance}). The eccentricity dependence (Fig.~\ref{fig:ecc_ob}), shows an asymmetry between the dominant jet (in the winter hemisphere) response of the short (NH summer) and long (SH summer) seasons. In the long season, the jet strength changes in a non-monotonic form with eccentricity. The jet strength increases in low eccentricity values (up to $\varepsilon\approx0.3$) and decreases for higher values. This response is consistent in both obliquity values. In contrast, in the short season, the jet strength strictly increases with eccentricity. The difference between the two seasons is a result of the temporal separation between the maximum and minimum temperature with increasing eccentricity, which occurs mainly during the long season (Fig.~\ref{fig:ecc_ob}). The jet strength dependence on perihelion is different. In this case, the long season (NH summer) jet decreases with perihelion, and for the short season (SH summer) jet, there is a general increase with perihelion. This response is a combination of the time delay between the minimum and maximum temperature, and the variations in solar constant during the seasonal cycle (Fig.~\ref{fig:ecc_per}). For example, in the $\Pi=90$ case (third column in Fig.~\ref{fig:ecc_per}), although the maximum and minimum align temporally, the meridional temperature gradient is weaker in the long season (NH summer), because the planet is farther from the host star. However, this alignment maintains the strong jet in the short season, compensating the warming in the winter hemisphere compared to other simulations.

\begin{figure}[htb!]
\centering
\includegraphics[width=\columnwidth]{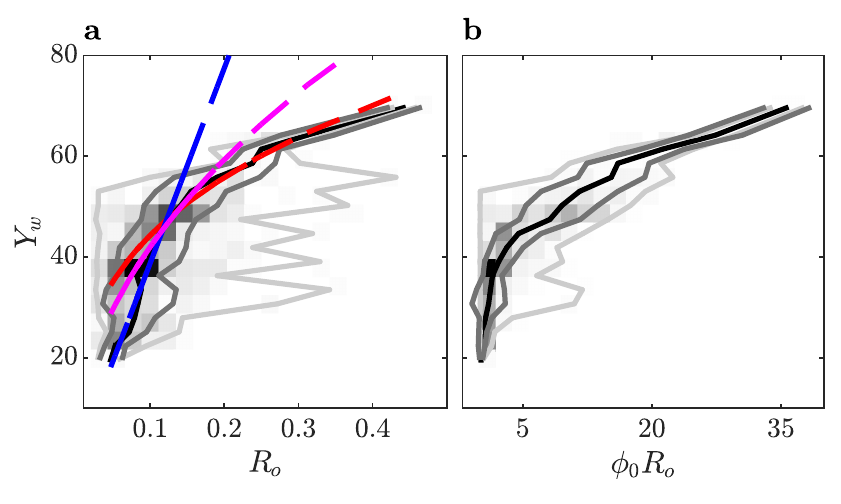}
\caption{Hadley cell width (in latitude degrees) as a function of the thermal Rossby number, $R_o$ (a), and the product of the thermal Rossby number with the latitude of maximum surface temperature, $\phi_0 R_o$ (b), during the seasonal cycle of all the simulations (shaded colors are occurrences, darker colors denote more abundant occurrences). Black line represents the bin average of the $Y_w$, dark gray lines are the standard deviations and the light gray lines confine all the points. The blue, magenta and red lines in (a) are for the scaling of $R_o$, $R_o^{1/2}$ and $R_o^{1/3}$, respectively. \label{fig:thermal}}
\end{figure}

The mean meridional circulation during the seasonal cycle is dominated mainly by a winter cell. This means that during the majority of the year, the circulation is composed of one cross-equatorial cell, with air rising in the summer hemisphere and descending in the winter hemisphere, and the transition seasons are relatively short (Figs.~\ref{fig:ecc_ob} and \ref{fig:ecc_per}). Note that, similar to the temperature response, there is also an asymmetry between the time periods of the circulation for each solstice season, with usually one shorter season with generally stronger circulation. The stronger circulation also occurs when the maximum temperature is at its most poleward position, and when it is closest to perihelion; this period generally also has higher $\Delta_T$ values. This correlation between the strength of the circulation, $\Delta_T$, and  the latitude of maximum temperature, $\phi_0$, is in agreement with axisymmetric arguments \citep{lindzen1988, guendelman2018, guendelman2019}. Similar arguments are given for the width of the circulation, $Y_w=|\phi_a-\phi_d|$, where $\phi_a$ is the latitude of the ascending branch and $\phi_d$ is the latitude of the descending branch; $Y_w$ is the width of the Hadley cell. According to the axisymmetric theory, the width of the circulation increases as the latitude of maximum temperature ($\phi_0$) goes poleward and as the thermal Rossby number (Eq. \ref{eq:thermal_rossby}) increases, which in this case can correspond to an increase in $\Delta_T$ \citep{guendelman2018}. That being said, there is a constraint on the circulation width (more specifically, the ascending branch of the circulation) that arises from axisymmetric considerations \citep{faulk_2017, guendelman2018, hill2019, singh2019}, where for planets with a low thermal Rossby number, $R_o$ (Eq.~\ref{eq:thermal_rossby}), the ascending branch will remain in midlatitudes, even if the maximum temperature is at the pole \citep{faulk_2017, guendelman2018}. Although these arguments are derived for a perpetual solstice case, and thus assumes fast adjustment to the radiative forcing, an assumption that is not necessarily accurate for this case, there is still a clear correlation between $Y_w$ and $R_o$, which becomes more pronounced (less spread) if the effect of $\phi_0$ is also taken into account (Fig.~\ref{fig:thermal}). The spread of the $Y_w$ in Figure~\ref{fig:thermal} can have several reasons: First, these are not a perpetual cases, meaning that the seasonal cycle is important. Second, to calculate $R_o$, we estimated the different parameters ($\Delta_T, \ H$) using the model output, in contrast to the original theory, where these parameters are input parameters of the model \citep{lindzen1988}.

Although the temperature response seems to follow the seasonal solar forcing, there is a need to examine the details of the seasonal cycle response. For example, looking at the cell width dependence on $R_o$, there seems to be a transition in the scaling from a linear response to a more complex power law (Fig.~\ref{fig:thermal}). The axisymmetric prediction is that the width of the circulation will follow $R_o^{1/2}$ for a perpetual equinox \citep{held1980} and $R_o^{1/3}$ for the perpetual solstice case \citep{caballero2008}. Then, we can assume that the different scaling in Figure \ref{fig:thermal} is the result of the seasonal cycle transition from a perpetual equinox scaling to perpetual solstice during the seasonal cycle. Alternatively, it is possible that this regime transition is not within the axisymmetric scaling, but rather from an eddy-mediated equinox circulation to an axisymmetric solstice circulation, similar to the transition suggested by \citet{bordoni2010}. A more detailed analysis of the dynamical response to the orbital configuration is needed, and is left for a future work.

Determining the climate on a tilted planet in an eccentric orbit is a complex problem that depends on various parameters. In addition to the dependence on the orbital parameters $\gamma$, $\varepsilon$, and $\Pi$, the climate strongly depends on parameters that control the atmospheric response, mainly ones that relate to the radiative timescale. The orbital period, atmospheric mass, and surface heat capacity are examples of important parameters that influence the resulting climate. Due to the variation in the timescales of the radiative changes during an eccentric orbit, the radiative timescale can significantly alter the climate response. The importance of the radiative timescale is also illustrated in section \ref{sec:zero_obliq}, where there is a qualitative difference in the climate response between short and long orbital periods. Examining the dependence of the climate response also on the radiative timescale or orbital period can also help illuminate the detailed seasonal cycle response of the temperature and circulation.

In contrast to the zero obliquity case, where the perpetual equinox dependence on the solar constant can be considered as the extreme limit of an infinitely long orbital period, there is no simple analog study for the non-zero obliquity case. The analog study in the non-zero obliquity case will be performing a perpetual study for each day in the seasonal cycle, and for each day, study its sensitivity to solar constant variations. Even if one does this type of study, its relevance will only be for very long orbital periods, as the timescale changes during the insolation seasonal cycle play an important role, and its importance increases with increasing eccentricity. Nonetheless, this type of study can act as a limit that can be compared with the seasonal cycle response and can help differentiate the seasonal transient effects.

\section{Conclusions} \label{sec:conc}
Studying the possible climate of exoplanets obliges us to think about the different possible orbital configurations and their effects on the climate. The simpler configurations are those in which the insolation is time independent, for example, perpetual equinox \citep[e.g.,][]{kaspi2015}, tidally locked \citep[e.g.,][]{merlis2010}, and perpetual reverse climates \citep{kang2019}. However, it is probable that a large number, if not the majority, of planets experience significant temporal variation in solar insolation. In this study, we focus on the effect of eccentricity on the diurnal mean climate for planets with zero and non-zero obliquity. It is important to note that changes in eccentricity and obliquity are not only between different planets, but also vary during the life time of a single planet that experiences Milankovich-like cycles \citep[e.g.,][]{spiegel_generalized_2010, way_effects_2017}, rendering this question even more relevant.

The insolation variations for a planet with zero obliquity in an eccentric orbit are equivalent to changes in the solar constant during the seasonal cycle. For this reason, studying the perpetual equinox response to variations in the solar constant is a good baseline for comparison with the seasonal cycle. Increasing the solar constant results in a trivial increase in temperature. However, the resulting climate can differ significantly among planets, depending on the moisture content of the atmosphere (Fig.~\ref{fig:solar1}). In the moist case, due to the non-linearity of the water vapor with temperature, the heat transport becomes more efficient (due to the latent heat flux, Fig.~\ref{fig:heat_flux}), resulting in a decrease in $\Delta_T$ with $S_0$, opposite to the dry case (Fig.~\ref{fig:solar1}a). These different responses have a strong effect on the atmospheric temperature profile and alter the dynamical response between moist and dry cases, emphasizing the role of moisture. This moisture-like dependence works for water, but also, in general, for any atmosphere with a condensible element, such as methane on Titan \citep[e.g.,][]{rannou2006, newman2016, mitchell2006, mitchell2008, mitchell2009, schneider2012, lora2014, lora2015}.

When seasonal variations are included, the radiative timescales become important, and the resulting climate will strongly depend on the ratio between the radiative timescale and the orbital period. For very short orbital periods, the main response of the atmosphere is to the annual mean forcing, where there is an increase in the mean flux with $\varepsilon$ \citep{bolmont_habitability_2016} (Fig.~\ref{fig:ecc_seas}c, e). As the orbital period becomes longer, there is a transition from a response dominated by a simple energy balance cases where other processes come into play, resulting in a response more similar to that of the perpetual case response to changes in $S_0$ (Figs.~\ref{fig:ecc_seas},~\ref{fig:ebm}). The temperature response to variations in eccentricity in a zero obliquity planet strongly depends on the orbital period and radiative timescale, specifically on the latitudinal structure of the radiative timescale (Fig.~\ref{fig:ebm}).

The temperature seasonal cycle couples with the seasonal cycle of the circulation. The Hadley circulation is affected by both the seasonal cycle of $\Delta_T$ and $\max(T_s)$, where it becomes stronger with steeper gradient and deeper with warmer temperature (Fig.\ref{fig:u_ecc}). The number of jets and their characteristics change for different orbital configurations and during the seasonal cycle. These changes result from the $\Delta_T$ seasonal cycle that constrains the eddy characteristics (Fig.~\ref{fig:u_ecc}).

Cases that combine changes in the obliquity and the eccentricity are more complex, as in addition to these two parameters the relative position between equinox and perihelion results in different insolation patterns (Fig.~\ref{fig:inso1}), making the solar forcing dependent on these three parameters (Fig.~\ref{fig:forcing}). In addition, the timescale over which the seasonal radiative changes occur also strongly depends on the orbital configuration. Specifically, at perihelion, the radiative forcing varies with a shorter timescale compared to the timescale at aphelion. These timescale differences widen with increasing eccentricity. As a result, the seasonal cycle of a tilted planet in an eccentric orbit is complex, with similar seasons having different climates in each hemisphere. Also, different seasons will have different timescales; for example, in the case of perihelion at equinox, there will be a fast transition between the two solstice seasons (Figs.~\ref{fig:ecc_ob}, \ref{fig:ecc_per}). 

These fast transitions in temperature result in the circulation also experiencing fast transitions during the seasonal cycle. Over most of the annual cycle, the Hadley circulation is composed of one cross-equatorial cell with air rising off the equator (with its direction depending on the specific season), with relatively short transition periods of two cells and air rising close to the equator. As in the temperature response, the period of each season is different, with the short season (close to perihelion) usually exhibiting stronger and wider circulation (Fig.~\ref{fig:thermal}). A good constraint on the circulation response is the thermal Rossby number and the latitude of maximum temperature, where poleward $\phi_0$ and higher $R_o$ will generally mean a stronger and wider circulation (Figs.~\ref{fig:ecc_ob}-\ref{fig:thermal}).

Although the general circulation response seems to follow the insolation, which puts a strong constraint on the circulation response, the details of the seasonal cycle can be complex \citep{merlis_hadley_2013-1}. A more detailed examination of the seasonal cycle in all the different orbital configurations is needed in order to better understand the climate on tilted planets in an eccentric orbit. It is possible that considering parameters such as the orbital period, and ones that relate to the surface and atmospheric radiative timescale, will reveal a qualitative difference in the atmospheric response. 

It has already been established that understanding the atmospheric dynamics is crucial to understand observations from exoplanets \citep[e.g.,][for the case of the equatorial winds advecting the subsolar hotspot in tidally locked exoplanets]{showman2002, knutson2007}. Here we show that the orbital configuration of a planet can alter the climate significantly. In addition, during the seasonal cycle the dynamics change considerably, emphasizing the importance of the seasonal cycle, in a large number (if not majority) of exoplanets. Although characterizing atmospheres of Earth-like planets orbiting Sun-like stars might not be within the upcoming missions' detection limit, planets in an eccentric orbit might be easier for characterization at different positions relative to their host star. This study shows the effect of the seasonal cycle on the atmospheric dynamics, that in turn can influence observations. This may motivate future studies on how a seasonal cycle can influence observations, and how to infer from these observations planetary characteristics \citep[for example, deduce from thermal curves information about the planets' obliquity, e.g.,][]{ rauscher2017models, ohno2019atmospheres2, adams_aquaplanet_2019}.

%In this study, we have shown the complexity of the seasonal response to variations in the orbital configuration. This complexity emphasizes the importance of understanding the seasonal cycle, as in the presence of a seasonal cycle, the climate differs substantially from the perpetual climate or the annual mean climate. It is reasonable to assume that many of the observed exoplanets have a wide variety of orbital configurations. These new and future understandings of the climate dependence on orbital configuration will help to advance the understanding of climate dynamics and might inspire future exoplanetary observations, as for the early theory superrotation in hot-Jupiters \citep{showman2002, knutson2007}.

\acknowledgments
This research was supported by the Israeli Space Agency and the Helen Kimmel Center for Planetary Science at the Weizmann Institute of Science. 

\bibliography{eccentricity}{}
\bibliographystyle{aasjournal}



\end{document}